\documentclass[aps,pra,twocolumn,a4paper,showpacs,superscriptaddress,floatfix,10pt]{revtex4}
\pdfoutput=1
\usepackage{graphicx}
\usepackage{amsmath}
\usepackage{epsfig}
\usepackage{helvet}
\usepackage{amssymb}
\usepackage{epstopdf}

\newcommand{\ket}[1]{\mbox{$| #1 \rangle$}}
\newcommand{\tr}{\mbox{tr}}
\newcommand{\proj}[1]{\mbox{$|#1\rangle \!\langle #1 |$}}
\newcommand{\matr}[1]{\left[\begin{matrix} #1 \end{matrix}\right]}

\def\OO{\mathbf{O}}

\def\AA{\mathbf{A}}
\def\QQ{\mathbf{Q}}
\def\GGamma{\mathbf{\Gamma}}
\def\II{\mathbf{I}}
\def\MM{\mathbf{M}}

\def\PP{\mathbf{P}}

\def\WW{\mathbf{W}}
\def\UU{\mathbf{U}}
\def\RR{\mathbf{R}}
\def\sss{s_A^{\mbox{\tiny ff}}}

\def\LLambda{\mathbf{\Lambda}}

\begin{document}

\title{Entanglement contour}

\author{Yangang Chen}
\affiliation{Perimeter Institute for Theoretical Physics, Waterloo, Ontario, N2L 2Y5, Canada}
\author{Guifre Vidal}
\affiliation{Perimeter Institute for Theoretical Physics, Waterloo, Ontario, N2L 2Y5, Canada}

\begin{abstract}
In the context of characterizing the structure of quantum entanglement in many-body systems, we introduce the entanglement contour, a tool to identify which real-space degrees of freedom contribute, and how much, to the entanglement of a region $A$ with the rest of the system $B$. The entanglement contour provides a complementary, more refined approach to characterizing entanglement than just considering the entanglement entropy between $A$ and $B$, with several concrete advantages. We illustrate this in the context of ground states and quantum quenches in fermionic quadratic systems. For instance, in a quantum critical system in $D=1$ spatial dimensions, the entanglement contour allows us to determine the central charge of the underlying conformal field theory from just a single partition of the system into regions $A$ and $B$, (using the entanglement entropy for the same task requires considering several partitions). In $D \geq 2$ dimensions, the entanglement contour can distinguish between gapped and gapless phases that obey a same boundary law for entanglement entropy. During a local or global quantum quench, the time-dependent contour provides a detailed account of the dynamics of entanglement, including propagating entanglement waves, which offers a microscopic explanation of the behavior of the entanglement entropy as a function of time.
\end{abstract}


\maketitle
\tableofcontents

\section{Introduction}

Over the last ten years, many-body entanglement has become the subject of intense theoretical investigations \cite{Boundary,QCrit,Logarithm}. There are important reasons to study quantum entanglement in many-body systems. On the one hand, the scaling of entanglement can be used as a tool to detect and characterize emergent quantum phenomena, such as quantum criticality \cite{QCrit} and topological order \cite{TO}. On the other hand, understanding the structure of many-body entanglement leads to efficient tensor network descriptions of many-body wave-functions \cite{TN}. Tensor networks can in turn be used to efficiently simulate many-body systems, and to classify the possible phases of quantum matter \cite{Classification}.

\subsection{A boundary law for ground state entanglement}

Most studies of many-body entanglement are based on characterizing bipartite entanglement. For that purpose, the many-body system is virtually divided into two regions, $A$ and $B$, and a measure of entanglement is used to quantify the entanglement between these regions. A prominent choice of measure of entanglement is the von Neumann entropy $S(A)$ of part $A$,
\begin{equation}
	S(A) \equiv -\tr(\rho^{A} \log_2 (\rho^{A})),
\end{equation}
where $\rho^{A} \equiv \tr_{B} \left( \proj{\Psi^{AB}} \right)$ is the density matrix for part $A$, and where we assumed that the whole system is in a pure state $\ket{\Psi^{AB}}$.

Let us specialize to the case where $\ket{\Psi^{AB}}$ is the ground state of a local Hamiltonian on a lattice in $D$ spatial dimensions. A central observation is that the entanglement between of a region $A$ of size $L^D$ and the (much larger) rest $B$ of the lattice is then often proportional to the size $|\sigma(A)|$ of the boundary $\sigma(A)$ of region $A$,
\begin{equation}
	S(A) \approx |\sigma(A)| \approx L^{D-1}. \label{eq:boundary}
\end{equation}
Thus, ground state entanglement typically obeys a \textit{boundary law} \cite{Boundary}, as opposed to the \textit{bulk law} $S(A) \approx |A| \approx L^D$ obeyed by generic states in the many-body Hilbert space. The boundary law of Eq. \ref{eq:boundary} is observed in the ground state of \textit{gapped} local Hamiltonian in arbitrary dimension $D$, as well as in some \textit{gapless} systems in $D>1$ dimensions. Instead, in gapless systems in $D=1$ dimensions, as well as in certain gapless systems in $D>1$ dimensions (namely systems with a Fermi surface of dimension $D-1$), ground state entanglement displays a logarithmic correction to the boundary law \cite{Logarithm},
\begin{equation}
	S(A) \approx |\sigma(A)|\log_2\left(|\sigma(A)|\right) \approx L^{D-1}\log_2 (L).
\label{eq:logarithm}
\end{equation}

\subsection{The contour of entanglement entropy}

At an intuitive level, the boundary law of Eq. \ref{eq:boundary} is understood as resulting from entanglement that involves degrees of freedom located near the boundary between regions $A$ and $B$. Also intuitively, the logarithmic correction of Eq. \ref{eq:logarithm} is argued to have its origin in contributions to entanglement from degrees of freedom that are further away from the boundary between $A$ and $B$.

The goal of this paper is to propose a simple formalism capable of testing the above intuitions, both qualitatively and quantitatively. Given the entanglement $S(A)$ between regions $A$ and $B$, we investigate the possibility of introducing a function $s_A$, the \textit{entanglement contour}, that assigns a real number $s_A(i)\geq 0$ to each lattice site $i$ contained in region $A$ such that the sum of $s_A(i)$ over all the sites $i\in A$ is equal to the entanglement entropy $S(A)$,
\begin{equation} \label{eq:aaa}
	S(A) = \sum_{i\in A} s_A(i),
\end{equation}
and that aims to quantifying how much the degrees of freedom in site $i$ participate in/contribute to the entanglement between $A$ and $B$. For instance, Fig. \ref{fig:1Dgapped} shows the entanglement contour for in gapped 1D systems (with open and periodic boundary conditions).

We emphasize at the outset that the entanglement contour $s_A(i)$ is \textit{not} equivalent to the von Neumann entropy $S(i) \equiv -\tr \left(\rho^{(i)} \log_2 \left(\rho^{(i)} \right) \right)$ of the reduced density matrix $\rho^{(i)}$ at site $i$. Notice that, indeed, the von Neumann entropy of individual sites in region $A$ is \textit{not} additive in the presence of correlations between the sites, and therefore generically
\begin{equation}
	S(A) \neq \sum_{i\in A} S(i),
\end{equation}
whereas the entanglement contour $s_A(i)$ is required to fulfill Eq. \ref{eq:aaa}. Relatedly, when site $i$ is only entangled with neighboring sites contained within region $A$, and it is thus uncorrelated with region $B$, the entanglement contour $s_A(i)$ will be required to vanish, whereas the one-site von Neumann entropy $S(i)$ still takes a non-zero value due to the presence of local entanglement within region $A$. 
 
\subsection{Structure of the paper}

This paper is divided into several sections. In section \ref{sect:contour} we introduce the requirements the entanglement contour $s_A(i)$ should satisfy. These requirements may be fulfilled by several functions, and therefore may not uniquely determine the entanglement contour. In Section \ref{sect:free} we then propose a specific function $\sss(i)$ within the free fermion formalism, and prove that it fulfills all the required constraints. Then the rest of the paper is devoted to exploring the proposed entanglement contour $\sss$ for free fermions in a number of different settings. First, in section \ref{sect:1D} we consider the ground state of quadratic fermionic Hamiltonians in one spatial dimension. There, the entanglement contour is seen to discriminate between gapped and gapless systems, and to display a universal profile at quantum criticality, from which one can extract the central charge of the underlying conformal field theory. In Section \ref{sect:2D} we then consider ground states of quadratic fermionic Hamiltonians in two spatial dimensions, and find that the entanglement contour is sensitive to whether the Hamiltonian is gapped or gapless and, in the latter case, to the presence of a Fermi surface. In section \ref{sect:quench} we consider time evolution after local and global quantum quenches \cite{LocalQuench,GlobalQuench}, and show that the entanglement contour unveils a detailed dynamical structure of propagation of entanglement within region $A$. Finally, Section \ref{sect:discussion} contains some discussion and conclusions. Overall, our results show that the entanglement contour provides useful insights into the spatial structure of many-body entanglement, well beyond what is already available from the entanglement entropy $S(A)$ alone and similar to what is possible with other tools (see \textit{Previous work} below).

\subsection{Reader's guide}

Sections \ref{sect:contour} and \ref{sect:free} contain several technical details that may well be skipped when first reading this manuscript. The key point is that the formalism of fermionic gaussian states offers a natural definition of the entanglement contour (in terms of probabilistic average of contributions coming from the fermionic modes that diagonalize the reduced density matrix $\rho^A$), see Eq. \ref{eq:fermion_si}, which fulfills all the requirements.

Sections \ref{sect:1D}-\ref{sect:quench} offer an easy-to-read, rather descriptive list of simple applications, which both demonstrate with concrete examples the use of the entanglement contour as a finer characterization of entanglement in extended quantum systems and suggest other possible applications and extensions of the results presented here. After becoming familiar with those examples, the interested reader may then want to go back to sections \ref{sect:contour} and \ref{sect:free} to learn in more detail the proposed formalism.

\subsection{Previous work}

The locality of entanglement entropy in many-body ground states has been previously probed using a number of other tools, some of which we briefly review below. We notice that none of these other tools, albeit extremely useful in characterizing the structure of many-body entanglement, define a proper contour for entanglement entropy, in the sense of fulfilling the basic requirement of Eq. \ref{eq:aaa}, namely that when summed up over region $A$, the contour should lead to the entanglement entropy of region $A$.

The \textit{quantum mutual information} $I(X:Y) \equiv S(X)+S(Y)-S(XY)$ between regions $X$ and $Y$ of a many-body system (where $X\cup Y$ is itself only part of the system) provides a measure of correlations between the two regions. It has been used to obtain insights into the structure of entanglement in the ground state of 1D systems and 2D systems, as well as in local and global quenches \cite{MutualInfo,LocalQuench,GlobalQuench}. 

The \textit{entanglement negativity} $\mathcal{N}(X:Y)$ \cite{Negativity} is a measure of entanglement of mixed states that has similarly been used to identify which local degrees of freedom contribute to the entanglement between two regions.

In Ref. \cite{spectra}, the locality of the entanglement in gapped systems was explored by developing a boundary-link perturbation theory capable of reproducing at the structure of entanglement spectrum.

The \textit{density of entanglement} $n(x,y)\equiv \partial_x \partial_y S(x,y)$ was proposed in Ref. \cite{Takayanagi} for continuous systems as the derivative of the entanglement entropy $S(x,y)$, where $x,y$ are the left and right boundaries of region $A$. The entanglement density has been effectively used to study local and global quenches in continuous systems, providing similar insight to the one offered in this context by the entanglement contour.

Much closer to our proposal are the studies of the lowest single-particle eigenfunctions of the entanglement Hamiltonian of Refs. \cite{EH}. Indeed, the entanglement contour for free fermions $\sss$, as introduced in Sect. \ref{sect:free}, can be interpreted as a weighted average over entanglement contributions made by the single-particle eigenfunctions of the entanglement Hamiltonian. In those settings (e.g. gapped 1D systems) where only a few such eigenfunctions are relevant, their real-space profile provides similar insights as the entanglement contour.

\section{Entanglement contour}
\label{sect:contour}

In this section we introduce the notion of entanglement contour and list a number of properties that it should fulfill.

Recall first that the entanglement between $A$ and $B$ is naturally described by the coefficients $\{p_{\alpha}\}$ appearing in the Schmidt decomposition of the state $\ket{\Psi^{AB}}$,
\begin{equation}\label{eq:Schmidt}
    \ket{\Psi^{AB}} = \sum_{\alpha} \sqrt{p_{\alpha}} \ket{\Psi^A_{\alpha}}\otimes \ket{\Psi^B_{\alpha}}.
\end{equation}
These coefficients $\{p_{\alpha}\}$ correspond to the eigenvalues of the reduced density matrix $\rho^A$, whose spectral decomposition reads
\begin{equation}\label{eq:rhoA1}
    \rho^A = \sum_{\alpha} p_{\alpha} \proj{\Psi_{\alpha}^A}.
\end{equation}
They define a probability distribution, $p_\alpha \geq 0$, $\sum_{\alpha} p_\alpha =1$, in terms of which the von Neumann entropy $S(A)$ is
\begin{equation}\label{eq:Sp}
    S(A) = -\sum_{\alpha} p_{\alpha} \log_2 (p_{\alpha}).
\end{equation}
On the other hand, the Hilbert space $\mathbb{V}^A$ of region $A$ factorizes as the tensor product
\begin{equation}\label{eq:VA}
    \mathbb{V}^{A} = \bigotimes_{i\in A} \mathbb{V}^{(i)},
\end{equation}
where $\mathbb{V}^{(i)}$ describes the local Hilbert space of site $i$.

The reduced density matrix $\rho^A$ in Eq. \ref{eq:rhoA1} and the factorization of Eq. \ref{eq:VA} define two inequivalent structures within the vector space $\mathbb{V}^{A}$ of region $A$. The entanglement contour $s_A$ is a function from the set of sites $i\in A$ to the real numbers,
\begin{equation}\label{eq:contour}
    s_A: A \longrightarrow \mathbb{R},
\end{equation}
that attempts to relate these two structures, by distributing the von-Neumann entropy $S(A)$ of Eq. \ref{eq:Sp} among the sites $i\in A$. 

\subsection{Five requirements}

Let us list the constraints we require on the entanglement contour $s_A(i)$. The first two conditions are very simple:
\begin{enumerate}
    \item Positivity: $s_A(i) \geq 0$.
    \item Normalization: $\displaystyle\sum_{i\in A} s_A(i) = S(A)$.
\end{enumerate}
These constraints amount to defining a probability distribution $p_i \equiv s_A(i)/S(A)$ over the sites $i\in A$, with $p_i \geq 0$ and $\sum_i p_i =1$, such that $s_A(i) = p_i S(A)$. Notice that these two conditions alone, however, do not require $s_A$ to inform us about the spatial structure of entanglement in $A$ --- they only relate to the density matrix $\rho_A$ through its total von Neumann entropy $S(A)$. A first step towards taking the explicit form of $\rho_A$ into account is by further requiring the following condition.
\begin{enumerate}
    \setcounter{enumi}{2}
    \item Symmetry: if $T$ is a symmetry of $\rho^A$, that is $T \rho^A T^{\dagger} = \rho^A$, and $T$ exchanges site $i$ with site $j$, then $s_A(i)=s_A(j)$.
\end{enumerate}
This condition ensures that the entanglement contour is the same on two sites $i$ and $j$ of region $A$ that, as far as entanglement is concerned, play an equivalent role in region $A$. It uses the (possible) presence of a spatial symmetry, such as invariance under space reflection, or under discrete translations/rotations, to define an equivalence relation in the set of sites of region $A$, and requires that the entanglement contour be constant within each resulting equivalence class. Notice, however, that this condition does not tell us whether the entanglement contour should be large or small on a given site (or equivalence class of site). In particular, the three conditions above are satisfied by a canonical choice $s_A(i) = S(A)/|A|$, that is a flat entanglement contour over the $|A|$ sites contained in region $A$, which once more does not tell us anything about the spatial structure of the von Neumann entropy in $\rho^A$. 

The remaining conditions refer to subregions within region $A$, instead of referring to single sites. It is therefore convenient to (trivially) extend the definition of entanglement contour to a set $X$ of sites in region $A$, $X \subseteq A$, with vector space
\begin{equation}\label{eq:VX}
    \mathbb{V}^X = \bigotimes_{i\in X} \mathbb{V}^{(i)},
\end{equation}
as the sum of the contour over the sites in $X$, 
\begin{equation}\label{eq:X}
    s_A(X) \equiv \sum_{i\in X} s_A(i).
\end{equation}
It follows from this extension that for any two disjoint subsets $X_1, X_2 \subseteq A$,  with $X_1 \cap X_2 = \emptyset$, the contour is additive, 
\begin{equation}\label{eq:additivity}
s_A(X_1 \cup X_2) = s_A(X_1) + s_A(X_2).
\end{equation}
In particular, condition 2 can be now recast as $s_A(A) = S(A)$. Similarly, if $X_1, X_2 \subseteq A$,  are such that all the sites of $X_1$ are also contained in $X_2$, $X_1 \subseteq X_2$, then the contour must be larger on $X_2$ than on $X_1$ (monotonicity of $s_A(X)$),
\begin{equation}\label{eq:monotonicity}
    s_A(X_1) \leq s_A(X_2)~~~~~\mbox{if}~~X_1 \subseteq X_2.
\end{equation}

Our next condition refers to transformations that cannot change the value of $s_A(X)$ on a subset of sites $X\subseteq A$.
\begin{enumerate}
    \setcounter{enumi}{3}
    \item Invariance under local unitary transformations: if the state $\ket{\Psi'^{AB}}$ is obtained from the state $\ket{\Psi^{AB}}$ by means of a unitary transformation $U^{X}$ that acts on a subset $X \subseteq A$ of sites of region $A$, that is $\ket{\Psi'^{AB}} \equiv U^X \ket{\Psi^{AB}}$, then the entanglement contour $s_A(X)$ must be the same for state $\ket{\Psi^{AB}}$ and for state $\ket{\Psi'^{AB}}$.
\end{enumerate}
That is, the contribution of region $X$ to the entanglement between $A$ and $B$ is not affected by a redefinition of the sites or change of basis within region $X$. Notice that it follows that $U^X$ can also not change $s_A(\bar{X})$, where $\bar{X}\equiv A-X$ is the complement of $X$ in $A$.

To motivate our last condition, let us consider a state $\ket{\Psi_{AB}}$ that factorizes as the product
\begin{equation}\label{eq:factorization0}
    \ket{\Psi^{AB}} = \ket{\Psi^{X X_B}} \otimes \ket{\Psi^{\bar{X} \bar{X}_B}},
\end{equation}
where $X\subseteq A$ and $X_B\subseteq B$ are subsets of sites in regions $A$ and $B$, respectively, and $\bar{X}\subseteq A$ and $\bar{X}_B\subseteq B$ are their complements within $A$ and $B$, so that  
\begin{eqnarray} \label{eq:Omega0}
  \mathbb{V}^A &=& \mathbb{V}^{X} \otimes \mathbb{V}^{\bar{X}}, \\
  \mathbb{V}^B &=& \mathbb{V}^{X_B} \otimes \mathbb{V}^{\bar{X}_B}.
\end{eqnarray}
Notice that in this case the reduced density matrix $\rho^A$ factorizes as $\rho^A = \rho^X \otimes \rho^{\bar{X}}$ and the entanglement entropy is additive,
\begin{equation}\label{eq:additivity0}
S(A) = S(X) + S(\bar{X}).
\end{equation}
Since the entanglement entropy $S(X)$ of subregion $X$ is well-defined, we demand that the entanglement profile over $X$ be equal to it,
\begin{equation}
s_A(X) = S(X).
\end{equation}

Our last condition refers to a more general situation where, instead of obeying Eq. \ref{eq:factorization0}, the state $\ket{\Psi_{AB}}$ factorizes as the product
\begin{equation}\label{eq:factorization}
    \ket{\Psi^{AB}} = \ket{\Psi^{\Omega_A\Omega_B}} \otimes \ket{\Psi^{\bar{\Omega}_A\bar{\Omega}_B}},
\end{equation}
with respect to some decomposition of $\mathbb{V}^A$ and $\mathbb{V}^B$ as tensor products of factor spaces,
\begin{eqnarray} \label{eq:Omega}
  \mathbb{V}^A &=& \mathbb{V}^{\Omega_A} \otimes \mathbb{V}^{\bar{\Omega}_A}, \\
  \mathbb{V}^B &=& \mathbb{V}^{\Omega_B} \otimes \mathbb{V}^{\bar{\Omega}_B}.
\end{eqnarray}
We emphasize that these factor spaces may not correspond to subsets of sites in $A$ or $B$ as in Eq. \ref{eq:Omega0}. Let $S(\Omega_A)$ denote the entanglement entropy supported on the first factor space $\mathbb{V}^{\Omega_A}$ of $\mathbb{V}^A$, that is
\begin{eqnarray} \label{eq:Omega1}
    S(\Omega_A) &=& - \tr \left(\rho^{\Omega_A} \log_2(\rho^{\Omega_A})\right), \\
    \rho^{\Omega_A} &\equiv& \tr_{\Omega_B} \left( \proj{\Psi^{\Omega_A\Omega_B}} \right), \label{eq:Omega2}
\end{eqnarray}
and let $X\subseteq A$ be a subset of sites whose vector space $\mathbb{V}^{X}$ is completely contained in $\mathbb{V}^{\Omega_A}$, meaning that $\mathbb{V}^{\Omega_A}$ can be further decomposed as
\begin{equation}\label{eq:Omega3}
    \mathbb{V}^{\Omega_A} \cong \mathbb{V}^{X}\otimes\mathbb{V}^{X'}.
\end{equation}
Then we arrive to our fifth condition.

\begin{enumerate}
    \setcounter{enumi}{4}
    \item Upper bound: if a subregion $X \subseteq A$ is contained in a factor space $\Omega_A$ (Eqs. \ref{eq:Omega} and \ref{eq:Omega3}) then the entanglement contour of subregion $X$ cannot be larger than the entanglement entropy $S(\Omega_A)$ (Eq. \ref{eq:Omega1})
        \begin{equation}\label{eq:upperbound}
            s_A(X) \leq S(\Omega_A).
        \end{equation}
\end{enumerate}

This condition says that whenever we can ascribe a concrete value $S(\Omega_A)$ of the entanglement entropy to a factor space $\Omega_A$ within region $A$ (that is, whenever the state $\ket{\Psi^{AB}}$ factorizes as in \ref{eq:Omega}) then the entanglement contour has to be consistent with this fact, meaning that the contour $S(X)$ in any subregion $X$ contained in the factor space $\Omega_A$ is upper bounded by $S(\Omega_A)$.

Let us consider a particular case of condition 5. When a region $X \in A$ is not at all correlated with $B$, that is $\rho^{XB} = \rho^{X}\otimes \rho^B$, then it can be seen \cite{Factorization} that $X$ is contained in some factor space $\Omega_A$ such that the state $\ket{\Psi^{\Omega_A\Omega_B}}$ itself further factorizes as $\ket{\Psi^{\Omega_A}}\otimes\ket{\Psi^{\Omega_B}}$, so that Eq. \ref{eq:factorization} becomes 
\begin{equation}\label{eq:factorization2}
    \ket{\Psi^{AB}} = \left(\ket{\Psi^{\Omega_A}} \otimes \ket{\Psi^{\Omega_B}} \right) \otimes \ket{\Psi^{\bar{\Omega}_A\bar{\Omega}_B}},
\end{equation}
and $S(\Omega_A)=0$. Condition 5 then requires that $s_A(X) =0$, that is 
\begin{equation}\label{eq:unentangledX}
    \rho^{XB} = \rho^{X}\otimes \rho^B \Rightarrow s_A(X) = 0,
\end{equation}
reflecting the fact that a region $X\subseteq A$ that is not correlated with $B$ does not contribute at all to the entanglement between $A$ and $B$.
 
Finally, we point out that the upper bound in condition 5 can be alternatively announced as a lower bound. Let $Y\subseteq A$ be a subset of sites whose vector space $\mathbb{V}^{Y}$ completely contains $\mathbb{V}^{\Omega_A}$ in Eq. \ref{eq:Omega0}, meaning that $\mathbb{V}^{Y}$ can be further decomposed as
\begin{equation}\label{eq:Omega4}
\mathbb{V}^{Y} \cong \mathbb{V}^{\Omega_A}\otimes\mathbb{V}^{\Omega_A'}.
\end{equation}
\begin{enumerate}
    \item[5$'$.] Lower bound: The entanglement contour of subregion $Y$ is at least equal to the entanglement entropy $S(\Omega_A)$ in Eq. \ref{eq:Omega1},
\begin{equation}\label{eq:upperbound}
    s_A(Y) \geq S(\Omega_A).
\end{equation}
\end{enumerate}
[Indeed, this can be seen to follow from the normalization $s_A(A)=S(A)$ of Condition 2 and the upper bound of Condition 5, by exchanging $\Omega_A$ and $\bar{\Omega}_A$ and choosing $Y \equiv A - X$.]

Conditions 1-5 are not expected to completely determine the entanglement contour. In other words, there probably are inequivalent functions $s_A:A \rightarrow \mathbb{R}$ that conform to all the conditions above. It might well be possible to extend the above list with additional conditions.

\subsection{Other measures of entanglement}

We conclude this section by noting that in our construction, the von Neumann entropy $S(A)$ was chosen for the sake of concreteness. An entanglement contour can also be defined for \textit{any} measure of pure-state entanglement which is additive under tensor product (in the sense of Eq. \ref{eq:additivity0}), such as the Renyi entropy of index $n$, 
\begin{equation}
	S_n(A) \equiv \frac{1}{1-n} \log_2 \left( \tr\left( (\rho^{A})^{n} \right) \right),
\end{equation}
for $n\geq 0$, which reduces to the von Neumann entropy in the limit $n\rightarrow 1$. Thus, we could define a contour $s_{n,A}$ by requiring that it fulfills conditions 1-5 by replacing the entanglement entropy $S$ with the Renyi entropy $S_n$ in all the corresponding expressions.

\section{Entanglement contour for free fermions}
\label{sect:free}

In this section we introduce a specific function $\sss(i)$ using the free fermion formalism, and show that it fulfills conditions 1-5 for an entanglement contour. We start by minimally reviewing the free fermion formalism (see e.g. \cite{Kitaev}).

\subsection{N fermionic modes}

Let us consider a fermionic lattice system made of $N$ sites, as characterized by a set of $N$ fermionic annihilation operators $a_i$, $i=1,2,\cdots,N$, which fulfill
\begin{equation}\label{eq:anti}
    \{a_i^{\dagger}, a_j\} = \delta_{ij},~~~~~~~~~~~ \{a_i, a_j\} = 0.
\end{equation}
Here we will mostly use the formalism of Majorana modes. There are two Majorana operators $\psi_{i,1}$ and $\psi_{i,2}$ associated to each site $i$, defined by
\begin{equation}\label{eq:Majorana}
    \psi_{i,1} \equiv \frac{a_i + a_i^{\dagger}}{\sqrt{2}},~~~~~~\psi_{i,2} \equiv \frac{a_i - a_i^{\dagger}}{i\sqrt{2}},
\end{equation}
or, equivalently,
\begin{equation}
	a_i	= \frac{\psi_{i,1} + i\psi_{i,2}}{\sqrt{2}},	\hspace{5mm}
	a_i^\dagger =\frac{\psi_{i,1} - i\psi_{i,2}}{\sqrt{2}}.
\end{equation}
Eqs. \ref{eq:anti} and \ref{eq:Majorana} imply the following anti-commutation relation
\begin{equation}\label{eq:antiM}
    \{\psi_{i,\alpha}, \psi_{j,\alpha'}\} = \delta_{i,j} \delta_{\alpha,\alpha'}.
\end{equation}

\subsection{Free fermion formalism}

From now on we restrict our attention to the case where the $N$ fermionic modes are in a gaussian state. A gaussian fermionic state of $N$ modes is the ground state of some quadratic Hamiltonian $H$, which in full generality reads:
\begin{equation}
	H = \frac{i}{2} \sum_{i,j=1}^{N} \sum_{\alpha,\alpha'=1}^{2} \psi_{i,\alpha} \AA_{i,\alpha;j,\alpha'} \psi_{j,\alpha'},
\end{equation}
If we regard $(i,\alpha)$ as a single index with $i=1,...,N$, $\alpha=1,2$, then $\AA$ is a $2N \times 2N$ generic real, anti-symmetric matrix that satisfies $\AA_{i,\alpha;j,\alpha'} = -\AA_{j,\alpha';i,\alpha}$.
By Wick's theorem, the gaussian state is completely characterized in terms of a $2N \times 2N$ antisymmetrized correlation matrix $\GGamma$,
\begin{equation}
	\GGamma_{i,\alpha;j,\alpha'} = -i\langle[\psi_{i,\alpha},\psi_{j,\alpha'}] \rangle ~~~~~~~ i,j = 1,...,N, ~ \alpha,\alpha'=1,2
\end{equation}
where $(\GGamma_{i,\alpha;j,\alpha'})^{*} = \GGamma_{i,\alpha;j,\alpha'} = -\GGamma_{j,\alpha';i,\alpha}$.

Let $Q \in$ O$(2N)$ be the orthogonal matrix that block diagonalizes $A$,
\begin{equation}\label{eq:diagA}
    \AA = \QQ\left( \bigoplus_{k=1}^{N}  \matr{
	0& \varepsilon_k \\
	-\varepsilon_k & 0\\
	} \right) \QQ^{T}.
\end{equation}
Then $\GGamma$ has the form (see e.g. \cite{Kitaev}),
\begin{equation}\label{eq:diagGamma}
    \GGamma = \QQ\left( \bigoplus_{k=1}^{N}  \matr{
	0& 1 \\
	-1 & 0\\
	} \right) \QQ^{T}.
\end{equation}
The reduced density matrix $\rho^{A}$ for a subset $A$ of $L$ sites, with $L\leq N$, is completely specified by the restriction $\GGamma^A$ of the correlation matrix $\GGamma$ on $A$. To simplify the notation, we can always reorganize the $N$ sites of the system in such a way that region $A$ corresponds to the first $L$ sites. Then the correlation matrix $\GGamma^A$ on region $A$ reads
\begin{equation}\label{eq:GammaA}
\GGamma^A_{i,\alpha;j,\alpha'} \equiv \GGamma_{i,\alpha;j,\alpha'}, ~~~~~~~ i,j = 1,...,L, ~ \alpha,\alpha'=1,2.
\end{equation}

\subsection{Entanglement entropy}

In order to compute the von Neumann entropy of $\rho^A$, we first block diagonalize
$\GGamma^A_{i,\alpha;j,\alpha'}$
by means of an orthogonal matrix $\OO \in $ O$(2L)$,
\begin{equation}\label{eq:diagGammaA}
        \GGamma^A = \OO\LLambda \OO^{T}, ~~~\LLambda \equiv  \bigoplus_{k=1}^{L}  \matr{
	0& \nu_k \\
	-\nu_k & 0\\
	}
\end{equation}
where
\begin{equation}\label{eq:ortho}
    \sum_{i,\alpha} \OO_{i,\alpha;k,\beta}\OO_{i,\alpha;k',\beta'} = \delta_{k,k'}\delta_{\beta,\beta'}
\end{equation}
with $k,k'=1,\cdots,L$ and $\beta,\beta'=1,2$, and define the delocalized Majorana modes $\{\phi_{k,\beta}\}$ by
\begin{equation}\label{eq:MajoranaPhi}
    \phi_{k,\beta} \equiv \sum_{i=1}^{L} \sum_{\alpha=1}^{2} (\OO^T)_{k,\beta; i,\alpha} \psi_{i,\alpha}.
\end{equation}
We can invert this relation to obtain
\begin{equation}\label{eq:MajoranaPhi}
    \psi_{i,\alpha} \equiv \sum_{k=1}^{L} \sum_{\beta=1}^{2} \OO_{i,\alpha;k,\beta} \phi_{k,\beta}.
\end{equation}
Eq. \ref{eq:diagGammaA} tells us that the pair of modes $\phi_{k,1}$ and $\phi_{k,2}$ are correlated with each other, $\langle \phi_{k,1} \phi_{k,2} \rangle = i\nu_k/2$, but uncorrelated with the rest of modes. This means that, when expressed in some adequate basis, the density matrix $\rho_A$ decomposes as the tensor product of $L$ density matrices $\varrho^{(k)}$ of size $2\times 2$,
\begin{equation} \label{eq:rhoA}
	U \rho^A U^{\dagger}= \bigotimes_{k=1}^{L} \varrho^{(k)} = \bigotimes_{k=1}^L  \matr{
	\frac{1+\nu_k}{2}&  0 \\
	0 & \frac{1-\nu_k}{2}\\},
\end{equation}
where $U$ is some $2^L \times 2^L$ unitary transformation related to the $2L \times 2L$ orthogonal matrix $O$.
Therefore, the von Neumann entropy of part $A$ is the sum of $L$ contributions, each corresponding to a pair $\phi_{k,1}, \phi_{k,2}$ of delocalized Majorana modes,
\begin{equation}
	S(A) = \sum_{k=1}^{L} S(k),
\end{equation}
where
\begin{equation} \label{eq:Sk}
	S(k) \equiv -\left(
	  \frac{1+\nu_k}{2} \log_2 \frac{1+\nu_k}{2}
	+ \frac{1-\nu_k}{2} \log_2 \frac{1-\nu_k}{2}
	\right).
\end{equation}

For later reference, we note that the entanglement entropy can be compactly expressed as
\begin{equation}\label{eq:BlogB}
    S(A) = \tr \left(f(\GGamma^A)\right),
\end{equation}
where
\begin{equation}\label{eq:BlogB2}
    f(\GGamma^A) \equiv  -\left(\frac{\II + i\GGamma^A}{2} \right)  \log_2\left(\frac{\II + i\GGamma^A}{2} \right).
\end{equation}
Here, the $2L \times 2L$ Hermitian matrix $(\II+i\GGamma^A)/2$ is the correlation matrix $\langle \psi_{i,\alpha} \psi_{j,\alpha'} \rangle$,
\begin{equation}\label{eq:B}
    \left(\frac{\II + i\GGamma^A}{2} \right)_{i,\alpha;j,\alpha'} = \langle \psi_{i,\alpha} \psi_{j,\alpha'} \rangle.
\end{equation}
This matrix can be diagonalized by the product of the unitary matrix $\WW\in$ U$(2L)$
\begin{equation}\label{eq:V}
\WW \equiv \bigoplus_{k=1}^{L}  \frac{1}{\sqrt{2}} \matr{
1 & -i \\
	-i & 1\\
	}
\end{equation}
and the orthogonal matrix $\OO$ in Eq. \ref{eq:diagGammaA},
\begin{equation}\label{eq:B2}
    \frac{\II + i\GGamma^A}{2}  = \OO \WW\left( \bigoplus_{k=1}^{L}  \matr{
    \frac{1+\nu_k}{2} & 0 \\
	0 & \frac{1-\nu_k}{2}\\
	} \right) \WW^{\dagger} \OO^{T}.
\end{equation}
Eq. \ref{eq:BlogB} follows from this expression, using the fact that the matrices $\OO\WW$ and $\WW^{\dagger}\OO^{T}$ disappear in taking the trace.

\subsection{Entanglement contour}

We have seen that the entanglement entropy $S(A)$ is a sum of $L$ contributions from the delocalized modes
$\{\phi_{k,\beta}\}$,
and that the orthogonal transformation $\OO$ in Eq. \ref{eq:diagGammaA} connects these modes back to the real space modes
$\{\psi_{i,\alpha}\}$,
as highlighted in Eq. \ref{eq:MajoranaPhi}. For each site $i$, this orthogonal transformation can be used to define a probability $p_i(k)$
\begin{equation} \label{eq:fermion_pik}
	p_i(k) \equiv \frac{1}{2} \sum_{\alpha,\beta=1}^2 (\OO_{i,\alpha;k,\beta})^2
\end{equation}
which reflects the weight that the pair of delocalized Majorana modes $\phi_{k,1}$ and $\phi_{k,2}$ have in the Majorana modes $\psi_{i,1}$ and $\psi_{i,2}$ on that site, with
\begin{equation}\label{eq:p}
    p_i(k) \geq 0, ~~~~~~~ \sum_k p_i(k) = 1.
\end{equation}
This suggests that a natural way of defining the entanglement contour $\sss(i)$ is by attaching to site $i$ a sum of the entanglement contributions $S(k)$ in Eq. \ref{eq:Sk} from each pair $\phi_{k,1}, \phi_{k,2}$ of delocalized modes $k$, weighted by the corresponding probability $p_i(k)$ in Eq. \ref{eq:fermion_pik}, that is
\begin{equation} \label{eq:fermion_si}
	\sss(i) \equiv \sum_{k=1}^L p_i(k) S(k).
\end{equation}

We point out an alternative interpretation of Eq. \ref{eq:fermion_si}. Going back to the formalism of fermionic operators (as opposed to Majorana fermions), the probabilities $p_i(k)$ describe the spatial pattern of the single-particle eigenfunction of the reduced density matrix $\rho^A$ (or, of its logarithm, the entanglement Hamiltonian $H^A \equiv - \log (\rho^A)$). Accordingly, $\sss(i)$ is a weighted average (this time with the weight being given by the contribution $S(k)$ to the entanglement entropy) of the spatial patterns of all the single-particle eigenfunctions of $\rho^A$. In Ref. \cite{EH}, the spatial pattern $p_i(k)$ of the leading eigenfunctions of $\rho^A$ has been used to study the spatial structure of entanglement. The pattern $p_i(k)$ of an individual eigenfunction $k$ often displays oscillations as a function of the site $i$. Thanks to the weighted average in Eq. \ref{eq:fermion_si}, these oscillations are no longer present in the contour $\sss$.

Below we will show that this quantity fulfills conditions 1-5. For later reference, however, we first express $\sss(i)$ in Eq. \ref{eq:fermion_si} in a form analogous to Eq. \ref{eq:BlogB},
\begin{equation}\label{eq:siBlogB}
    \sss(i) = \tr \left( \MM^{(i)}  f(\GGamma^A) \right),
\end{equation}
where we have introduced the rank-2 projector
\begin{equation}\label{eq:Mi}
\MM^{(i)} \equiv \sum_{\alpha=1}^2 \proj{i,\alpha}.
\end{equation}
Indeed, some simple algebra involving Eq. \ref{eq:B2} shows that the right-hand-side of Eq. \ref{eq:siBlogB} amounts to
\begin{eqnarray}
    &&\sum_{\alpha,k,\beta,\beta'} \OO_{i,\alpha; k,\beta} \OO_{i,\alpha; k, \beta'} \times \label{eq:showsi1}\\
    &&\left(\delta_{\beta,\beta'} \frac{S(k)}{2} + (\delta_{\beta,1}\delta_{\beta',2} - \delta_{\beta,2}\delta_{\beta',1})g(k) \right) \label{eq:showsi2}\\
    &=& \sum_k \frac{1}{2}\left(\sum_{\alpha,\beta} (\OO_{i,\alpha;k,\beta})^2 \right) S(k) = \sss(i), \label{eq:showsi3}
\end{eqnarray}
where $g(k)$ is some function of $k$, and where in order to discard the second term in line \ref{eq:showsi2} we have used that, under the exchange of indices $\beta \leftrightarrow \beta'$, $\OO_{i,\alpha; k,\beta} \OO_{i,\alpha; k \beta'}$ is symmetric, whereas $\delta_{\beta,1}\delta_{\beta',2} - \delta_{\beta,2}\delta_{\beta',1}$ is antisymmetric.

\subsection{Proof of properties 1-5}

Next we show that $\sss(i)$ as defined in Eq. \ref{eq:fermion_si} fulfills conditions 1-5 of section \ref{sect:contour}.
\begin{enumerate}
    \item Positivity: $\sss(i) \geq 0$.
\end{enumerate}
Proof: This property follows from the fact that $\sss(i)$ in Eq. \ref{eq:fermion_si} is defined in terms of a sum of products of non-negative quantities $p_i(k)\geq 0$ and $S(k)\geq 0$.

\begin{enumerate}
    \setcounter{enumi}{1}
    \item Normalization: $\displaystyle\sum_{i\in A} \sss(i) = S(A)$.
\end{enumerate}
Proof: This results from a chain of simple equalities,
\begin{eqnarray}
\sum_i \sss(i) &=& \sum_{ik} p_i(k) S(k) = \sum_k S(k) \left(\sum_i p_i(k)\right) \nonumber \\
 &=& \sum_k S(k) = S(A), \label{eq:ProofCond2}
\end{eqnarray}
where in the third equality we have used $\sum_i p_i(k) = 1$, which follows from the orthogonal character of matrix $\OO$ in Eq. \ref{eq:fermion_pik}, see Eq. \ref{eq:ortho}.

\begin{enumerate}
    \setcounter{enumi}{2}
    \item Symmetry: if $T$ is a symmetry of $\rho^A$, that is $T \rho^A T^{\dagger} = \rho^A$, and $T$ exchanges site $i$ with site $j$, then $\sss(i)=\sss(j)$.
\end{enumerate}
Proof: Let $\PP \in $ SO$(2L)$ be the matrix that implements the symmetry transformation among the $2L$ Majorana modes, so that
\begin{eqnarray}\label{eq:SymGammaA}
    \PP^T \left( \GGamma^A \right) \PP &=& \GGamma^A,\\
    \PP \left( \MM^{(i)} \right) \PP^{T} &=& \MM^{(j)}. \label{eq:Mi2}
\end{eqnarray}
Notice that Eq. \ref{eq:SymGammaA} implies that
\begin{eqnarray}
    \PP^Tf(\GGamma^A)\PP = f(\GGamma^A).
\end{eqnarray}
It then follows that
\begin{eqnarray}\label{eq:symmProof}
    \sss(i) &=& -\tr \left( \MM^{(i)}   f(\GGamma^A) \right) \nonumber\\
    &=& -\tr \left( \MM^{(i)}  \left(\PP^T f(\GGamma^A) \PP\right) \right) \nonumber \\
    &=& -\tr \left( \left(\PP \MM^{(i)} \PP^T\right) f(\GGamma^A) \right) \nonumber\\
    &=& -\tr \left( \MM^{(j)}  f(\GGamma^A) \right)  = \sss(j).\nonumber
\end{eqnarray}

We emphasize that the orthogonal transformation $\PP$ does not need to merely map $\psi_{i,\alpha}$ into $\psi_{j,\alpha}$, but may also include an orthogonal transformation of modes within each site $i$. Indeed, Eq. \ref{eq:Mi2} only demands that the modes $\psi_{i,1}$ and $\psi_{i,2}$ be mapped into linear combinations of the modes $\psi_{j,1}$ and $\psi_{j,2}$.
\begin{enumerate}
    \setcounter{enumi}{3}
    \item Invariance under local unitary transformations: if the state $\ket{\Psi'^{AB}}$ is obtained from the state $\ket{\Psi^{AB}}$ by means of a unitary transformation $U^{X}$ that acts on a subset $X \subseteq A$ of sites of region $A$, that is $\ket{\Psi'^{AB}} \equiv U^X \ket{\Psi^{AB}}$, then the entanglement contour $s_A(X)$ must be the same for state $\ket{\Psi^{AB}}$ and for state $\ket{\Psi'^{AB}}$.
\end{enumerate}
Proof: Let $\OO^X \in$ SO$(2L^X)$ (where $L^X$ denotes the number of sites in region $X$) be the matrix that implements the unitary transformation $U^X$, so that
\begin{equation}\label{eq:Aprime}
    \GGamma'^{A} = \OO^{X}\GGamma^A \left( \OO^{X}\right)^{T},
\end{equation}
and let $\MM^X$ be a projector corresponding to the $2L^X$ modes in $X$,
\begin{equation}\label{eq:MX}
    \MM^X \equiv \sum_{i\in X} \MM^{(i)},
\end{equation}
which fulfill $[\MM^{X},\OO^{X}]=0$.
Since $\sss(X) \equiv \sum_{i \in X} \sss(i)$,
\begin{eqnarray}\label{eq:cond4}
    \sss(X) &=& \sum_{i\in X} \tr \left( \MM^{(i)}  f(\GGamma^A) \right) \\
    &=& \tr \left( \MM^{X}  f\left(\GGamma^A\right) \right) \\
    &=& \tr \left(\MM^{X}  f\left(\GGamma^A\right) \left( \OO^{X} \right)^T\OO^{X}\right) \\
    &=& \tr \left(\MM^{X}  \OO^{X} f\left(\GGamma^A\right)\left( \OO^{X} \right)^T\right) \\
    &=& \tr \left(\MM^{X}  f\left(\OO^{X}\GGamma^A\left( \OO^{X} \right)^T\right)\right) \\
    &=& \tr \left(\MM^{X}\left( f\left(\GGamma'^A \right) \right) \right) = {\sss}'(X),
\end{eqnarray}
where we have used that, for any $\RR\in$ SO$(2L)$, $\RR f(\GGamma^A)\RR^{T} = f(\RR \GGamma^A\RR^{T})$. This proves condition 4.
\begin{enumerate}
    \setcounter{enumi}{4}
    \item Upper bound: if a subregion $X \subseteq A$ is contained in the factor space $\Omega_A$ of Eqs. \ref{eq:factorization}, \ref{eq:Omega0}, and \ref{eq:Omega3}, then the entanglement contour of subregion $X$ cannot be larger than the entanglement entropy $S(\Omega_A)$ in Eq. \ref{eq:Omega1},
        \begin{equation}\label{eq:upperbound}
            \sss(X) \leq S(\Omega_A).
        \end{equation}
\end{enumerate}

Proof: Let us assume that region $X\subseteq A$ consists of the first $L^X$ sites in region $A$ (possibly, after permuting the sites within region $A$), and let $\bar{X}$ denote the rest of sites in $A$, that is $\bar{X} = A - X$. We can divide the real-space majorana modes into two groups:
\begin{equation}
	\begin{split}
	&\psi^{X} = \{\psi_{i,\alpha} \; | i\in X \},
	\\
	&\psi^{\bar{X}} = \{\psi_{i,\alpha} \; | i \in \bar{X}\},
	\end{split}
\end{equation}
and map them into another two groups of modes,
\begin{equation} \label{eq:psiP}
	\begin{split}
	&\eta^{\Omega_A} = \{\eta_{p,\gamma} \; | \; p \in \Omega_A\},
	\\
	&\eta^{\bar{\Omega}_A} = \{\eta_{p,\gamma} \; | \; p \in \bar{\Omega}_A\},
	\end{split}
\end{equation}
such that the set of real-space modes $\psi^{X}$ is completely contained in the set $\eta^{\Omega_A}$, meaning that
\begin{equation}
	\matr{
	\eta^{\Omega_A} &
	\eta^{\bar{\Omega}_A}
	}
	=
	\matr{
	\psi^{X} &
	\psi^{\bar{X}}
	}
	\matr{
	\mathbf{U}_{11} & 0 \\
	\mathbf{U}_{21} & \mathbf{U}_{22} \\
	}.
\end{equation}
Here $\UU_{11}$ is a rectangular matrix of dimensions $2L^{X} \times 2L^{\Omega_A}$, where $2L^{\Omega_A}$ is the number of Majorana modes in $\eta^{\Omega_A}$, and $L^{X} \leq L^{\Omega_A}$. In addition, by construction
\begin{equation}\label{eq:projX}
    \matr{
    \UU_{11}^T \UU_{11} & \\
    & 0
    }
    = \MM^{X},
\end{equation}
where $\MM^X$ is a projector onto the first $2L^X$ rows and columns of $\GGamma^{A}$.
The factorization $\ket{\Psi^{AB}} = \ket{\Psi^{\Omega_A\Omega_B}} \otimes \ket{\Psi^{\bar{\Omega}_A\bar{\Omega}_B}}$ in Eq. \ref{eq:factorization} implies the factorization $\rho^A = \rho^{\Omega_A}\otimes \rho^{\bar{\Omega}_A}$ of the density matrix in region $A$. This means that the modes $\eta^{\Omega_A}$ are not correlated with the modes $\eta^{\bar{\Omega}_A}$, so that when we express the correlation matrix $\GGamma^A$ in the basis of the modes $\eta_P$, we obtain a block diagonal form,
\begin{equation}
	\matr{
	\mathbf{U}^T_{11} & \mathbf{U}^T_{21} \\
	0 & \mathbf{U}_{22}^T \\
	}
	\Gamma^A
	\matr{
	\mathbf{U}_{11} & 0 \\
	\mathbf{U}_{21} & \mathbf{U}_{22} \\
	}
	=
	\matr{
	\GGamma^{\Omega_A} & \\
	& \GGamma^{\bar{\Omega}_A} \\
	}
\end{equation}

Then the entanglement contour for subregion $X\subseteq A$ is given by
\begin{eqnarray}\label{eq:sXfinal}
    \sss(X) &=& \tr \left( \MM^X f(\GGamma^A)\right) \\
    &=& \tr \left( \MM^X  f(\GGamma^{\Omega_A}) \right) \\
    &\leq& \tr\left( f(\GGamma^{\Omega_A}) \right) = S(\Omega_A), \nonumber
\end{eqnarray}
where in the inequality we have used that the operator $f(\GGamma^{\Omega_A})$ is non-negative defined.

\section{Ground states in one dimension}
\label{sect:1D}

In this section we investigate the entanglement contour $\sss$ for fermionic gaussian states, as defined in Eq. \ref{eq:fermion_si}, on the ground state of quadratic fermionic lattice Hamiltonians in $D=1$ spatial dimensions.

We consider the Hamiltonian
\begin{eqnarray} \nonumber
	H &=& -i \sum_{(i,j)}
	[ (1+\gamma) \psi_{i,1}\psi_{j,2}
	+ (1-\gamma) \psi_{j,1}\psi_{i,2} ] \\
	&+& i\mu \sum_i \psi_{i,1}\psi_{i,2},\label{eq:Hamiltonian}
\end{eqnarray}
where $(i,j)$ indicate that sites $i$ and $j$ are nearest neighbors. Using fermionic operators, this Hamiltonian reads
\begin{eqnarray}
	H &=& -\sum_{(i,j)}
	[ a^\dagger_i a_j + a^\dagger_j a_i
	+ \gamma (a_i a_j + a^\dagger_j a^\dagger_i) ] \\
	&+& \mu \sum_i a^\dagger_i a_i \label{eq:H}
\end{eqnarray}
Assuming periodic boundary conditions, Fourier and Bogoliubov transformations can be used to take this Hamiltonian into the diagonal form
\begin{equation}
	H = \sum_k E_k \alpha^\dagger_k \alpha_k,
\end{equation}
where $\alpha_k$ is a fermionic annihilation operator and the spectrum $E(k)$ of single-particle energies reads

\begin{equation}
	E(k)
	= \sqrt{
	\left(\mu-2 \cos k \right)^2
	+ \left(2\gamma \sin k \right)^2}.
\end{equation}
Direct inspection then reveals that Hamiltonian $H$ is gapless (that is, there exists at least one momentum label $k$ such that $E(k)=0$) if (1) $|\mu|<2, \gamma = 0$; or (2) $|\mu|=2, \forall \gamma$. Otherwise, $H$ has a finite energy gap in the thermodynamic limit.

\subsection{Gapped Hamiltonians}

\begin{figure}[h]
\begin{center}
\includegraphics[scale=0.5]{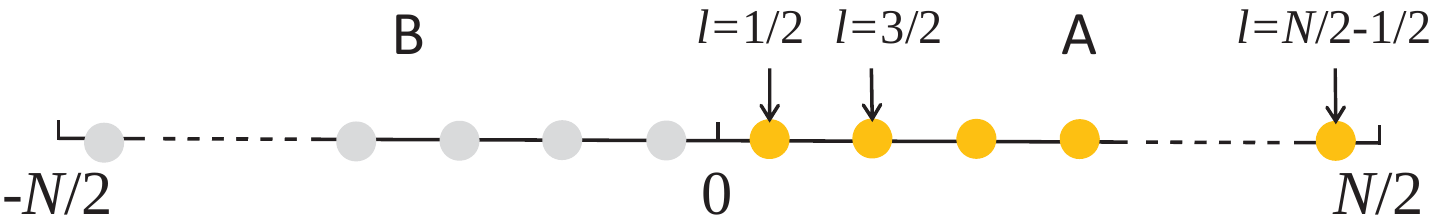}
\caption{
\label{fig:1Dchain1}
Lattice in $D=1$ dimensions made of $N$ sites and with open boundary conditions. The $N$ sites are at location $l = -\frac{N}{2}+\frac{1}{2},-\frac{N}{2}+\frac{3}{2}, ..., \frac{N}{2}-\frac{1}{2}$. Region $A$ contains $L = \frac{N}{2}$ sites, at location: $l = \frac{1}{2},\frac{3}{2}, ..., \frac{N}{2}-\frac{1}{2}$.
}
\end{center}
\end{figure}

Let us consider a gapped system, and let us focus first on the simplest possible setting: a chain with open boundary conditions, where region $A$, made of $L$ sites, corresponds to half of the chain of $N=2L$ sites. More specifically, the chain has sites at locations $l= -\frac{N}{2} + \frac{1}{2}, -\frac{N}{2}+\frac{3}{2}, \cdots, \frac{N}{2}-\frac{1}{2}$, and region $A$ corresponds to the $L$ sites at locations $l= \frac{1}{2}, \frac{3}{2}, \cdots, L-\frac{1}{2}$, see Fig. \ref{fig:1Dchain1}.

Recall that when the Hamiltonian is gapped, the entanglement of its ground state obeys a boundary law, Eq. \ref{eq:boundary}, which in $D=1$ dimensions translates into the saturation of $S(A)$ to a constant $S^{(\infty)}$ when the size $L$ of region $A$ becomes larger than the correlation length $\xi$ in the ground state of $H$,
\begin{equation}\label{eq:Ssaturation}
    S(A) \approx S^{(\infty)},~~~~~~L \gg \xi~~~\mbox{(OBC)}.
\end{equation}
What type of entanglement contour do we then expect? A plausible guess is that the contour $s_A(l)$ should decay exponentially with the distance $l$ to the boundary of $A$, for distances $l$ larger than the correlation length $\xi$,
\begin{equation}\label{eq:ansatz}
    s_A(l) \approx \alpha e^{-l/\xi},~~~~~~~~~~l > \xi
\end{equation}
where $\alpha$ is some constant. An exponential decay is expected because the off-diagonal elements of the correlation matrix $\Gamma^A$ decay exponentially with the distance to the diagonal of $\Gamma^A$ (when the distance to the diagonal is larger than $\xi$), implying that all correlations in the ground state also decay exponentially with distance. For the sake of argument, let us assume that in the whole range $l \in [\frac{1}{2},L-\frac{1}{2}]$, the contour $s_A(l)$ is independent of $L$. Then summing $s_A(l)$ over all the sites in region $A$, we obtain
\begin{eqnarray}\label{eq:}
    S(A) &=& \sum_{l=\frac{1}{2}}^{L-\frac{1}{2}} s_A(l) \\ 
    &\approx&  \sum_{l=\frac{1}{2}}^\xi s_A(l) + \alpha \sum_{l=\xi+1}^{L-\frac{1}{2}} e^{-l/\xi}  \label{eq:step1}\\
    &=&  \sum_{l=\frac{1}{2}}^\xi s_A(l) + \alpha \sum_{l=\xi+1}^{\infty} e^{-l/\xi} + O(e^{-L/\xi}) \label{eq:step2}\\
    &=& S^{(\infty)} + O(e^{-L/\xi}),\label{eq:step3}
\end{eqnarray}
where in Eq. \ref{eq:step1} we have symbolically broken the sum into two contributions corresponding to $l\leq \xi$ and $l > \xi$, and replaced $s_A(l)$ with its exponential form for $l > \xi$ (Eq. \ref{eq:ansatz}). In addition, in Eq. \ref{eq:step2} we have changed the size of region $A$ from a finite number $L$ of sites to an infinite number ---thus relating $S(A)$ with the constant $S^{(\infty)}$ in Eq. \ref{eq:Ssaturation}--- by introducing only a correction $O(e^{-L/\xi})$ that is exponentially suppressed in $L/\xi$. Thus, assuming that $s_A(l)$ decays exponentially with $l$ is, up to the small correction $O(e^{-L/\xi})$ in Eq. \ref{eq:step3}, indeed consistent with the expected saturation of the entanglement entropy $S(A)$ to a constant $S^{(\infty)}$ as a function of $L$.

\begin{figure}[h]
\begin{center}
\includegraphics[scale=0.48]{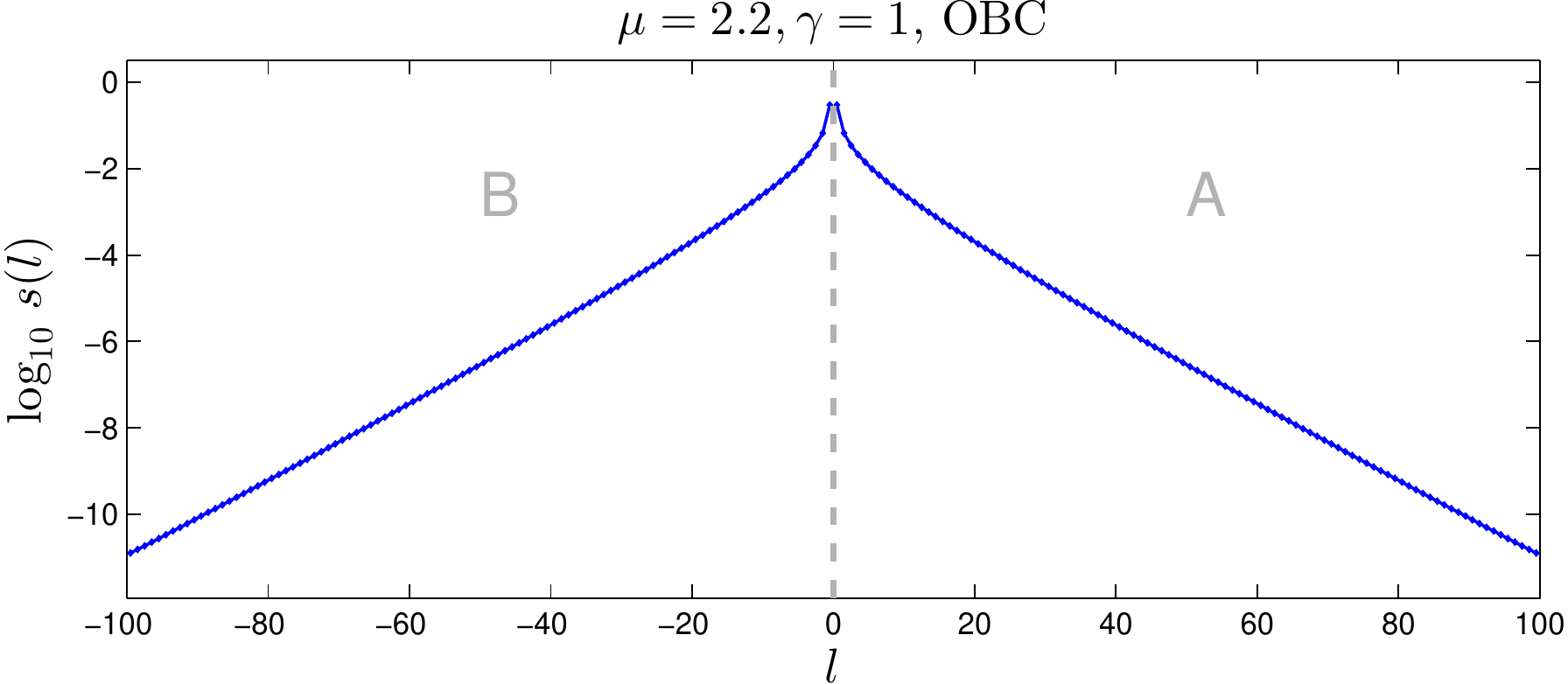}
\includegraphics[scale=0.48]{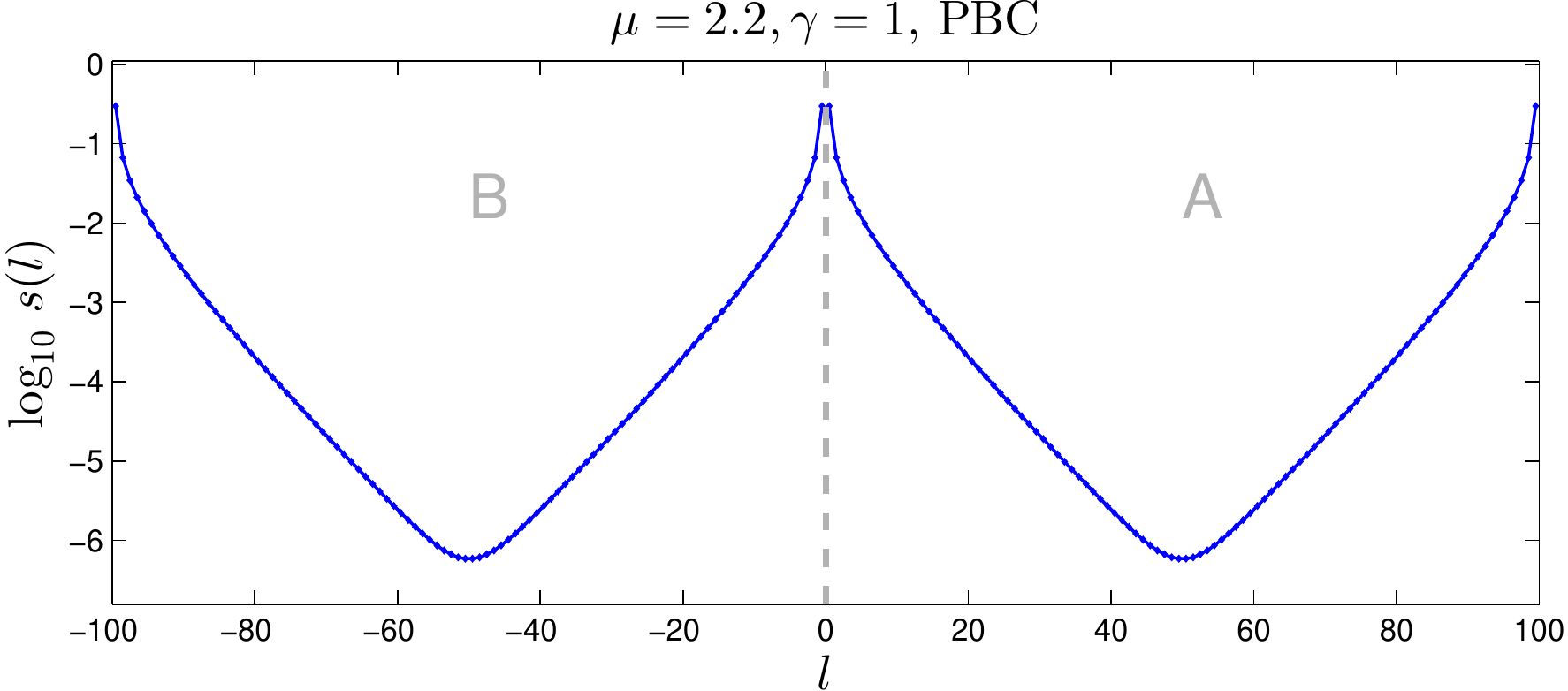}
\caption{
\label{fig:1Dgapped}
Entanglement contour for regions $A$ and $B$ in a gapped chain ($\mu = 2.2, \gamma = 1$). The chain contains $N=200$ sites and has open (top) and periodic (bottom) boundary conditions. Regions $A$ and $B$ have $L=100$ adjacent sites each. The contours decay exponentially with the distance to the closest boundary.
}
\end{center}
\end{figure}

Let us then evaluate the entanglement contour $\sss(l)$ and see if it conforms to Eq. \ref{eq:ansatz}. For that purpose, we choose $\mu=2.2$ and $\gamma=1$, for which the ground state has a correlation length $\xi$ of just a few lattice sites. The upper panel of Fig. \ref{fig:1Dgapped} shows the entanglement contour $\sss$ on region $A$ and $s_B^{\mbox{\tiny ff}}$ on region $B$, each made of $L=100$ sites, on a lattice with $N=200$ sites and open boundary conditions. The contour is seen to decay exponentially with the distance to the boundary between $A$ and $B$ for $l\gg \xi$, consistent with our guess above. Near the boundary ($l \leq \xi$), a different, less dramatic decay is observed, as discussed later.

The lower panel of Fig. \ref{fig:1Dgapped} shows the entanglement contour for the ground state of the same Hamiltonian but with periodic boundary conditions. In this case, the boundary of $A$ with $B$ consists of two points or cuts (at $l=0$ and $l=L$), and the entanglement entropy $S(A)$ of a large region $A$ is known to saturate to twice $S^{(\infty)}$,
\begin{equation}\label{eq:Ssaturation2}
    S(A) \approx 2 \times S^{(\infty)},~~~~~~L \gg \xi~~~\mbox{(PBC)}.
\end{equation}
We see that now the contour $\sss(l)$ is roughly the sum of two independent contributions, one per each boundary cut, that again decay exponentially with the distance to the relevant cut. This is fully compatible with Eq. \ref{eq:Ssaturation2}, where $S(A)$ can be understood to be the sum of two contributions, one for each boundary cut.

We conclude that, in a gapped system in $D=1$ dimensions, the entanglement contour $\sss$ allows us to formalize, through its exponential decay with the distance to the closest boundary of region $A$ (Eq. \ref{eq:ansatz}), the intuition that ground state entanglement involves essentially the degrees of freedom that are near the boundary.

Our conclusions are consistent with those obtained previously using other tools, such as mutual information \cite{MutualInfo}, study of the eigenfunctions of the entanglement Hamiltonian of free fermions \cite{EH}, or a perturbative study of the entanglement spectrum \cite{spectra}.

\subsection{Gapless Hamiltonians}

Let us now consider a gapless Hamiltonian. We start again with a chain of size $N=2L$, Fig. \ref{fig:1Dchain1}, with open boundary conditions. In this case the entanglement entropy $S(A)$ displays a logarithmic correction to the boundary law, Eq. \ref{eq:logarithm}, which for a large chain reads \cite{QCrit}
\begin{equation}\label{eq:SlogL1}
    S(A) = \frac{c}{6} \log_2 (L) + O(1),    ~~~~~L\gg 1,~~ \mbox{(OBC)}.
\end{equation}
Here $c$ is the central charge of the conformal field theory (CFT) that describes the quantum critical system at long distances.
Specifically, the Hamiltonian of Eq. \ref{eq:Hamiltonian} corresponds to a quantum critical system of spinless fermions described by a CFT with central charge $c=\frac{1}{2},1$, for
\begin{equation}\label{eq:ccc}
    c = \left\{ \begin{array}{cl}
          \frac{1}{2} & ~~~\mbox{for } |\mu|=2, \forall \gamma, \\
          &\\
          1 & ~~~\mbox{for } |\mu|<2, \gamma = 0.
        \end{array} \right.
\end{equation}

What contour should we expect for these systems? A natural guess is that this time the contour should decay as 
\begin{equation}\label{eq:power}
    s_A(l) \approx \frac{c}{6\ln 2}\cdot\frac{1}{l},
\end{equation}
because then
\begin{eqnarray}\label{eq:sumpower}
    S(A) &=& \sum_{l=\frac{1}{2}}^{L-\frac{1}{2}} s_A(l) = \frac{c}{6\ln 2} \sum_{l=\frac{1}{2}}^{L-\frac{1}{2}} \frac{1}{l}\\
    &\approx&  \frac{c}{6\ln 2} \int_{\frac{1}{2}}^{L-\frac{1}{2}} \frac{dx}{x} = \frac{c}{6} \log_2 (L) + O(1),
\end{eqnarray}
which agrees with eq. \ref{eq:SlogL1}.

\begin{figure}[h]
\begin{center}
\includegraphics[scale=0.48]{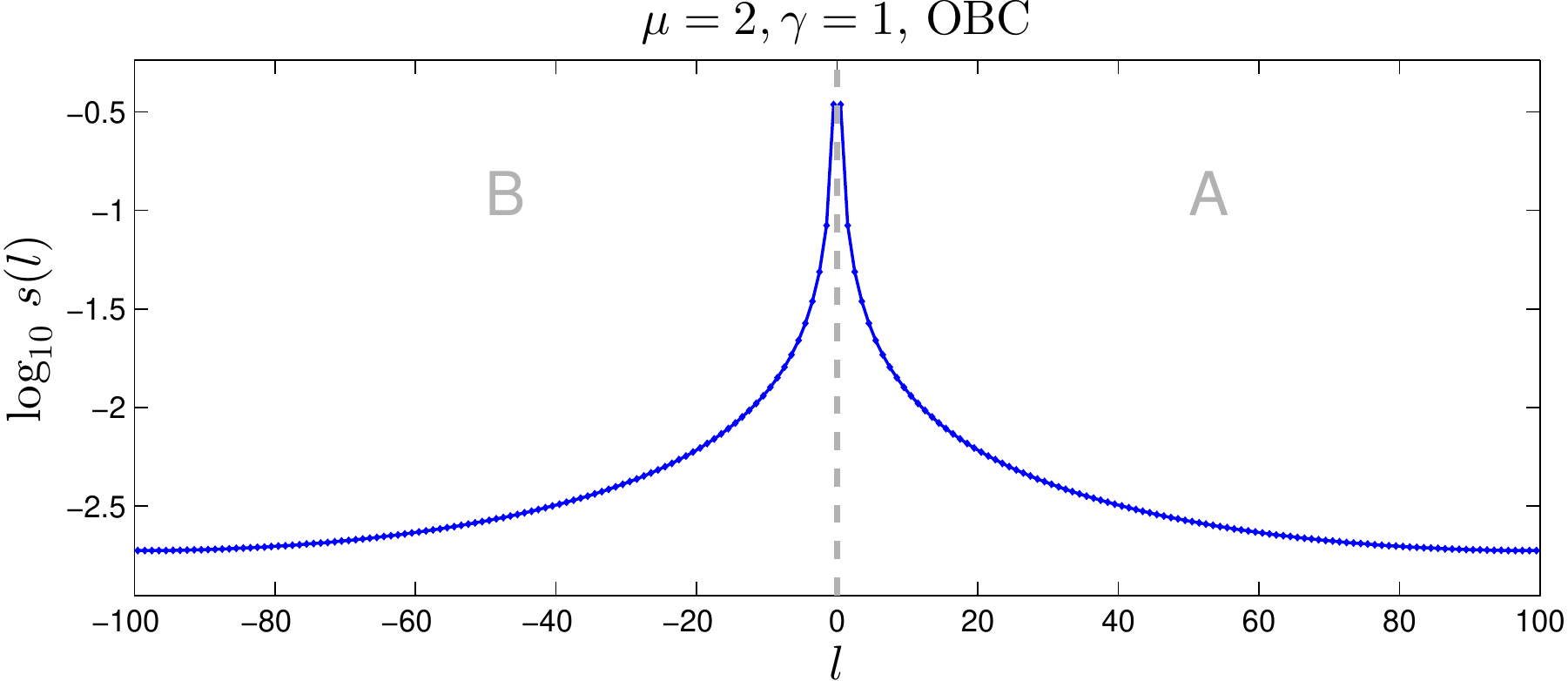}
\includegraphics[scale=0.48]{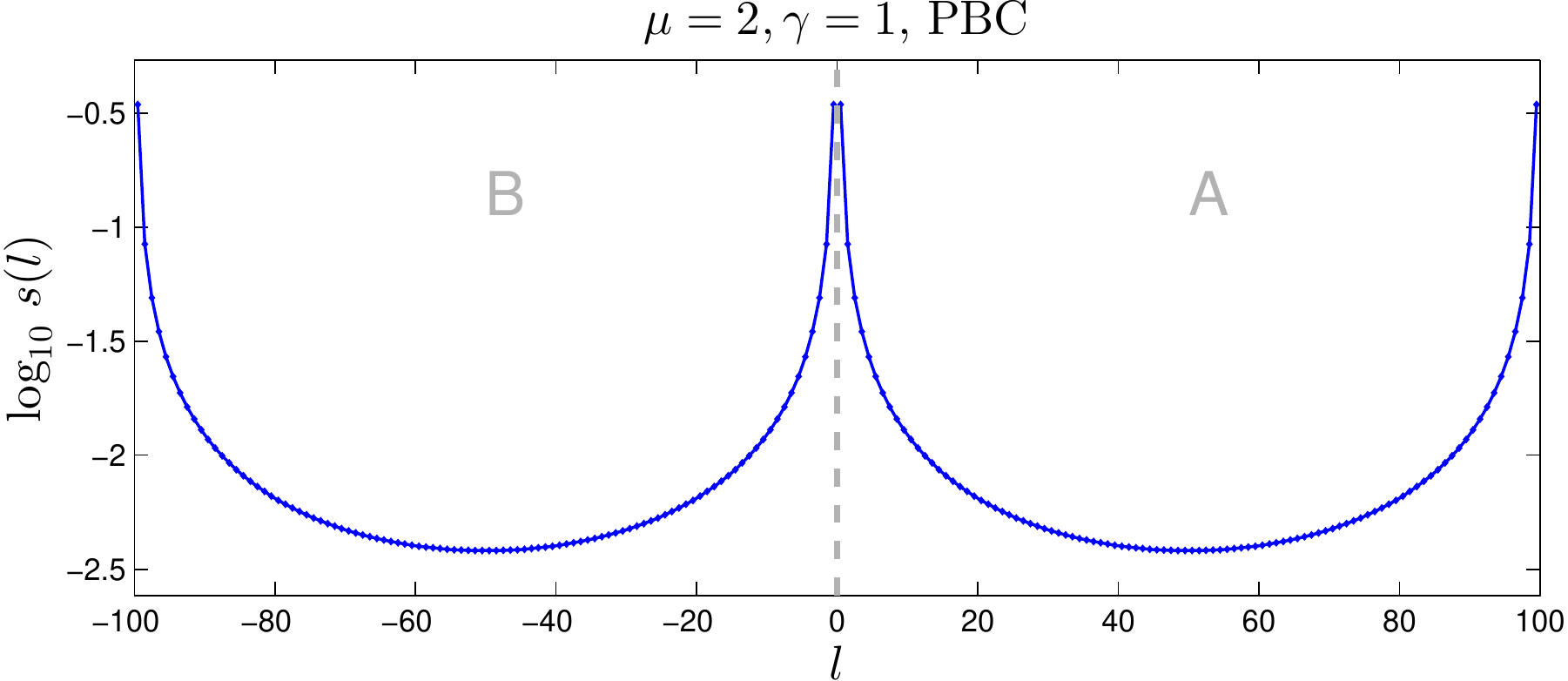}
\caption{
\label{fig:1Dgapless}
Entanglement contour for regions $A$ and $B$ in a gapless chain ($\mu = 2, \gamma = 1$). The chain contains $N=200$ sites and has open (top) and periodic (bottom) boundary conditions. Regions $A$ and $B$ have $L=100$ adjacent sites each. The contours decay roughly as a power law with the distance to the closest boundary.
}
\end{center}
\end{figure}

\begin{figure}[h]
\begin{center}
\includegraphics[scale=0.6]{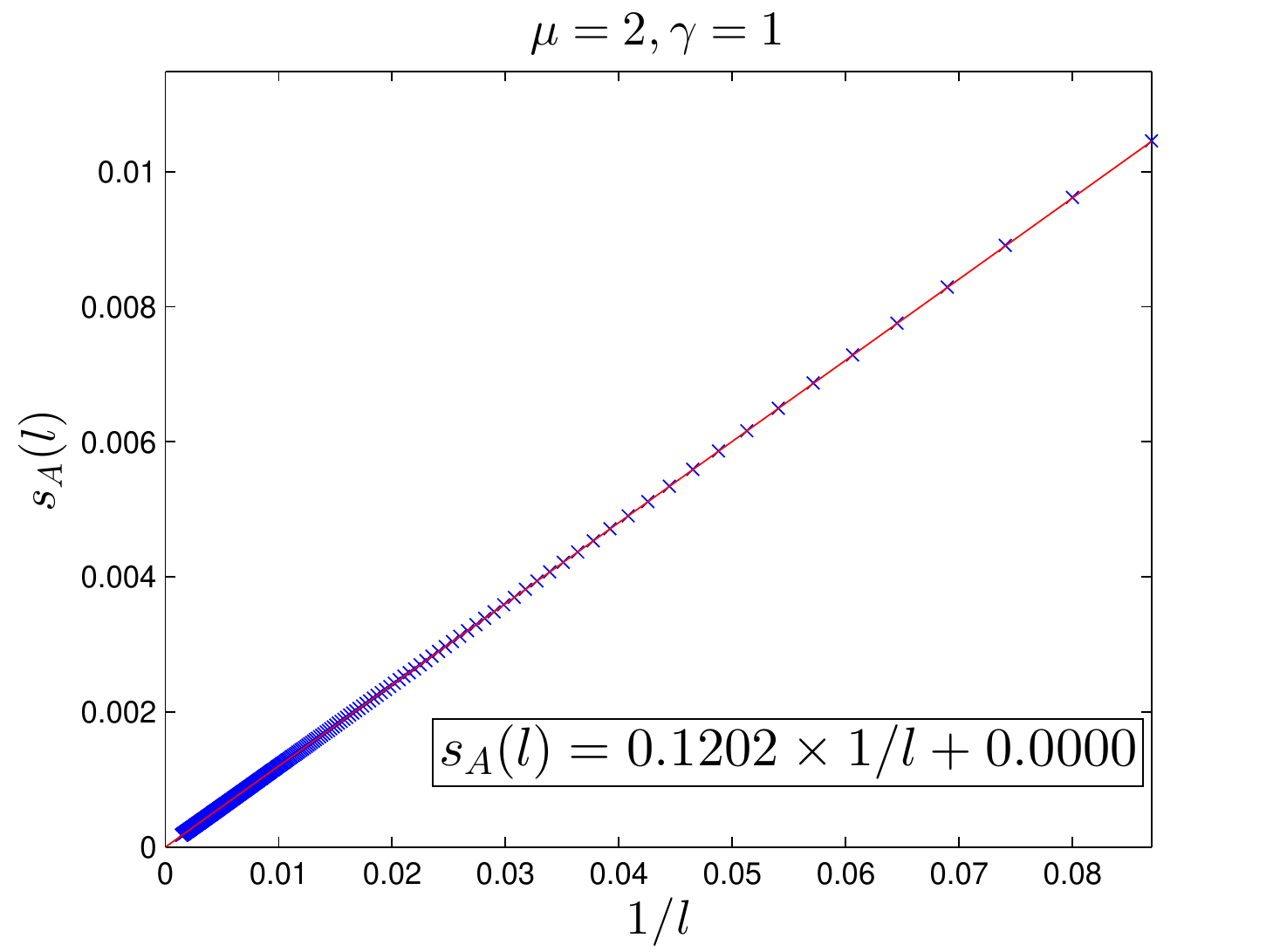}
\caption{
\label{fig:CFT_OBC}
Entanglement contour $s_A(l)$ as a function of $1/l$, for $(\mu = 2, \gamma = 1)$. The plot corresponds to a chain made of $N=5000$ sites with open boundary conditions, where region $A$ is made of $L=N/2=2500$ sites, see Fig. \ref{fig:1Dchain1}. We display the contour $s_A(l)$ for $l=11.5,12.5,...,699.5$, which reveal an accurate $1/l$ dependence, with a coefficient $0.1202 \approx \frac{1}{12\ln 2}$ that encodes the central charge of the corresponding CFT: $\frac{c}{6\ln 2} \approx \frac{1}{12\ln 2} \, \Rightarrow \, c \approx \frac{1}{2}$. The correlation coefficient of the linear fitting is $R = 1-7.4\times 10^{-7}$.}
\end{center}
\end{figure}

The upper panel of Fig. \ref{fig:1Dgapless} shows the entanglement contour $\sss(l)$ for the gapless chain $(\mu=2, \gamma =1)$. At distances $l$ from the boundary that are sufficiently large compared to $1$ but still small compared to the size of the region $A$, $1 \ll l \ll L$, $\sss(l)$ matches Eq. \ref{eq:power} very well, and can be used to extract the central charge $c=\frac{1}{2}$, see Fig. \ref{fig:CFT_OBC}. Thus, we can use the entanglement contour of the entanglement entropy of a \textit{single} region $A$ to estimate the central charge $c$, instead of having to use Eq. \ref{eq:SlogL1} to fit $S(A)$ for several regions $A$ with different size $L$ as it is commonly done.

The lower panel of Fig. \ref{fig:1Dgapless} corresponds to periodic boundary conditions. The entanglement entropy is known to scale as \cite{QCrit}
\begin{equation}\label{eq:SlogL2}
    S(A) = 2 \times \frac{c}{6} \log_2 (L) +O(1),    ~~~~~L\gg 1,~~ \mbox{(PBC)},
\end{equation}
that is, as twice its value for the chain with OBC, immediately suggesting that we will again find a contour with roughly two independent contributions of the form \ref{eq:power}, one for each of the two boundary cuts. This is indeed what Fig. \ref{fig:1Dgapless} shows.

We conclude that, in a gapless system in $D=1$ dimensions, the entanglement contour allows us to formalize the intuition that, again, the entanglement involves predominantly the degrees of freedom that are near their joint boundary. However, in contrast with what happens for gapped system, this time the degrees of freedom that are further away from the boundary also contribute to the entanglement entropy, as expressed in Eq. \ref{eq:power}.

A similar conclusion can be reached e.g. by studying the mutual information using a large number of subregions \cite{MutualInfo}. Notice that here we only had to compute the entanglement profile for a single region.

\subsection{Estimating the central charge from the entanglement contour of a region $A$ with two boundary cuts}

\begin{figure}[h]
\begin{center}
\includegraphics[scale=0.5]{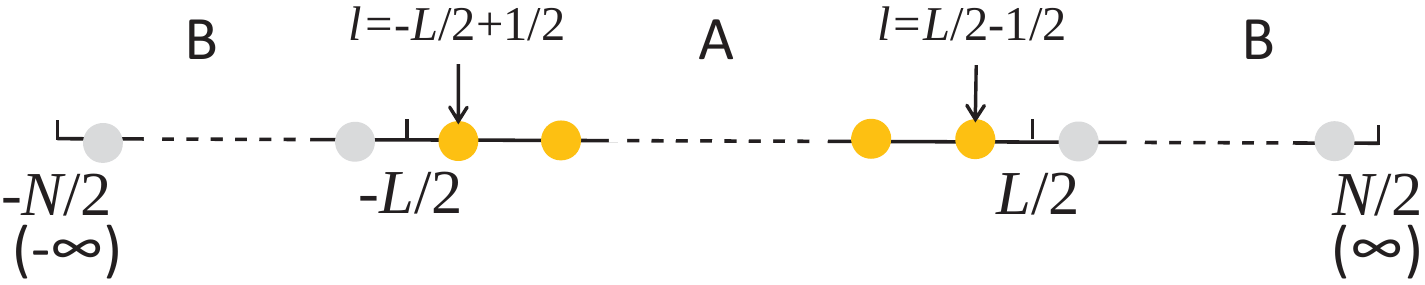}
\caption{
\label{fig:1Dchain2}
Lattice in $D=1$ dimensions made of $N$ sites and with periodic boundary conditions. We consider the thermodynamic limit, $N\to \infty$. Region $A$ contains a finite number $L$ of adjacent sites, at positions $l = -\frac{L}{2}+\frac{1}{2}, -\frac{L}{2}+\frac{3}{2}, ..., \frac{L}{2}-\frac{1}{2}$.
}
\end{center}
\end{figure}

Let us consider the special case of a finite region $A$ of size $L$ in an infinite system, $N\rightarrow \infty$, where the sites of $A$ are at positions $l=-\frac{L}{2}+\frac{1}{2}, -\frac{L}{2} + \frac{3}{2}, \cdots, \frac{L}{2} - \frac{1}{2}$, see Fig. \ref{fig:1Dchain2}.

\begin{figure}[h]
\begin{center}
\includegraphics[scale=0.6]{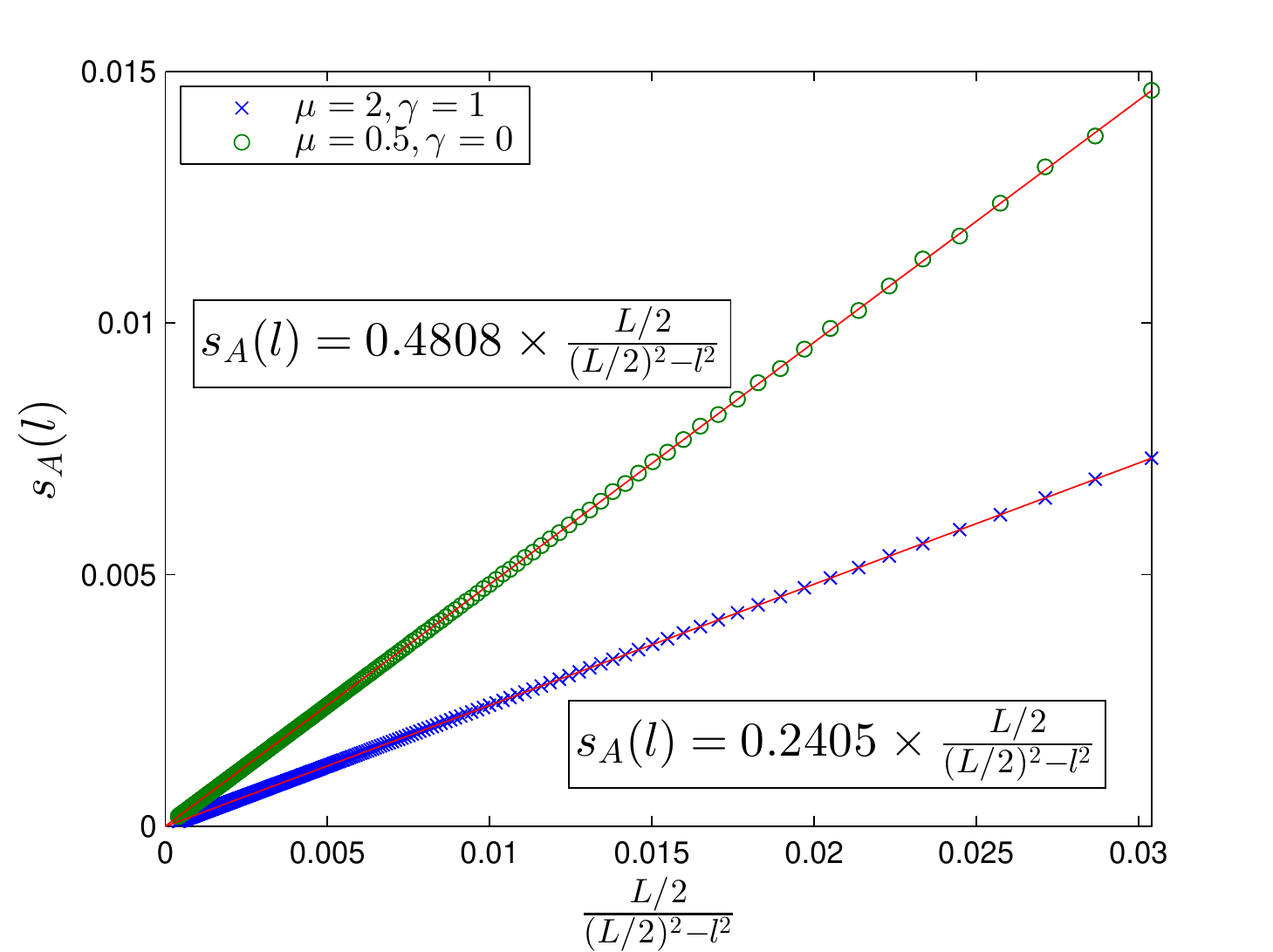}
\caption{
\label{fig:CFT_PBC}
Entanglement contour $s_A(l)$ as a function of $\frac{L/2}{(L/2-l)(L/2+l)}$. We consider a chain made of $N=\infty$ sites, with a region $A$ made of $L=5001$ sites, see Fig. \ref{fig:1Dchain2}. We plot the contour of the following points: $l = -2484,-2483,...,2484$.
\\
(Blue crosses): For ($\mu = 2, \gamma = 1$), the contour $s_A(l)$ is proportional to $\frac{L/2}{(L/2-l)(L/2+l)}$, with proportionality coefficient $0.2405 \approx \frac{1}{6\ln 2}$. We can estimate the central charge of the underlying CFT, obtaining $\frac{c}{3\ln 2}\approx\frac{1}{6\ln 2} \, \Rightarrow \, c \approx \frac{1}{2}$. The correlation coefficient is $R = 1-3.0\times 10^{-9}$.
\\
(Green circles): For ($\mu = 0.5, \gamma = 0$), the proportionality coefficient is $0.4808 \approx \frac{1}{3\ln 2}$, so that $\frac{c}{3\ln 2}\approx \frac{1}{3\ln 2} \, \Rightarrow \, c \approx 1$. The correlation coefficient is $R = 1-3.1\times 10^{-6}$.
}
\end{center}
\end{figure}

Fig. \ref{fig:CFT_PBC} shows that for $L\gg 1$ the entanglement contour is very well approximated by the simple expression
\begin{equation} \label{eq:CFTProfile}
	s_A(l)
	= \frac{c}{3\ln 2} \frac{\frac{L}{2}}{(\frac{L}{2}-l)(\frac{L}{2}+l)}.
\end{equation}
This form of the contour coincides with an exact CFT calculation by Robert Myers \cite{Myers}. In the large $L$ limit, we can replace the sum over sites by an integral and indeed obtain
\begin{equation}
	\begin{split}
	&\sum_{l=-\frac{L}{2}+\frac{1}{2}}^{\frac{L}{2}-\frac{1}{2}} s_A(l)
	= \frac{c}{3\ln 2} \sum_{l=-\frac{L}{2}+\frac{1}{2}}^{\frac{L}{2}-\frac{1}{2}} \frac{\frac{L}{2}}{(\frac{L}{2}-l)(\frac{L}{2}+l)}
	\\
	=& \frac{c}{3\ln 2} \int_{-1+\frac{1}{L}}^{1-\frac{1}{L}} dx \; \frac{1}{1-x^2}
	= \frac{c}{3} \log_2(L) + O(1),
	\end{split}
\end{equation}
which coincides with Eq. \ref{eq:SlogL2}. Specifically, Fig. \ref{fig:CFT_PBC} shows the entanglement contour for $(\mu=2, \gamma=1)$, corresponding to a CFT with central charge $c=\frac{1}{2}$, and $(\mu=0.5, \gamma=0)$, corresponding to a CFT with central charge $c=1$. It is clear from the figure that the entanglement contour of a region $A$ with two boundary cuts can again be used to obtain precise estimates of the central charge.

Once more, the central charge can be obtained from the entanglement entropy or the mutual information by studying a large number of subregions \cite{MutualInfo}. Here we only had to compute the entanglement profile for a single region.

\subsection{Approaching a critical point}

\begin{figure}[h]
\begin{center}
\includegraphics[scale=0.5]{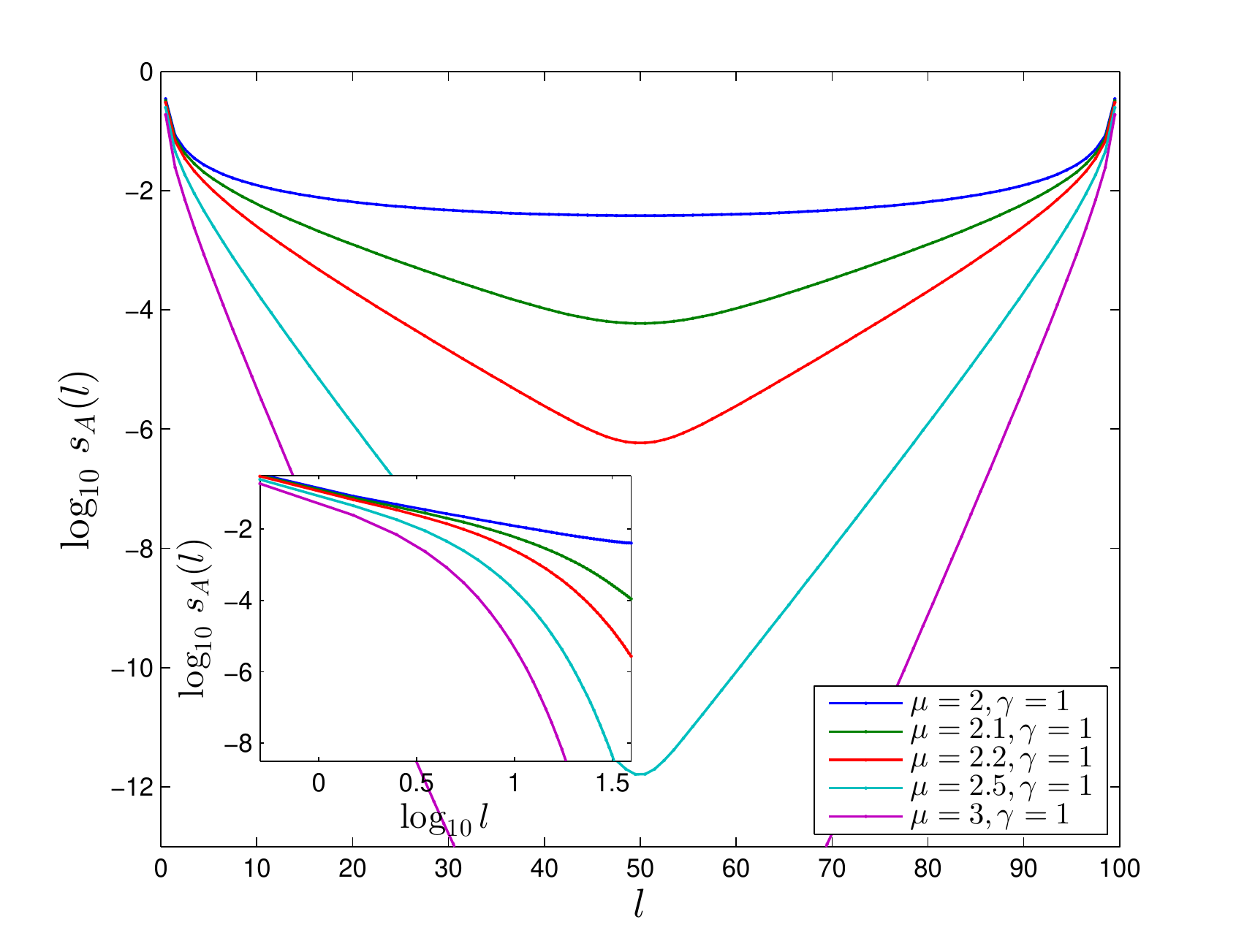}
\caption{
\label{fig:1Dgappedgapless}
Comparison, in a lattice of $N=200$ sites and with PBC, of the entanglement contour of a fixed region $A$ of $L=N/2=100$ sites (see Fig. \ref{fig:1Dchain1}) for different ground states, corresponding to a gapless Hamiltonian ($\mu = 2; \gamma = 1$), upper-most line,  and a sequence of gapped Hamiltonians ($\mu = 2.1,2.2,2.5,3; \gamma = 1$), rest of lines. One can see that as $\mu$ approaches the critical value $\mu=2$, the entanglement contour of the ground state of the gapped system approaches that of the gapless system. The inset shows the same contour but closer to the left boundary. The critical contour decays as a power law with the distance to the boundary, whereas each gapped contour decays exponentially for distances larger than the correlation length $\xi$. However, as the gap closes and the correlation length $\xi$ grows, the contour of the gap system resembles more and more that of the critical system.
}
\end{center}
\end{figure}

To conclude this section, it is instructive to investigate how the exponential contour of Eq. \ref{eq:ansatz} for a gapped system transforms into the power law contour of Eq. \ref{eq:power} for a gapless system. This is illustrated in Fig. \ref{fig:1Dgappedgapless}, which shows the entanglement contour in the setting of Fig. \ref{fig:1Dchain2} for a sequence of values of $(\mu,\gamma)$ in the Hamiltonian $H$, corresponding to gapped systems with decreasing energy gap. As the gap becomes smaller, the correlation length $\xi$ becomes larger, diverging for a gapless system. Roughly speaking, the entanglement contour is seen to follow a universal power-law for $l \ll \xi$, and the exponential scaling of Eq. \ref{eq:ansatz} for $\xi \ll l \ll L$, a behavior that can be approximately interpolated by
\begin{equation}\label{eq:yukawa}
    s_A(l)~ \propto ~~c~\frac{e^{-l/\xi}}{l},~~~~~~~~~~~~~~l \ll L/2.
\end{equation}

This expression can actually be derived from Eq. (22) in the third paper in Ref. \cite{EH} for some particular models.

\section{Ground states in two dimensions}
\label{sect:2D}

In this section we investigate the entanglement contour $\sss$ for fermionic gaussian states, as defined in Eq. \ref{eq:fermion_si}, on the ground state of quadratic fermionic lattice Hamiltonians in $D=2$ spatial dimensions.
For this purpose, we consider the Hamiltonian of Eqs. \ref{eq:Hamiltonian}-\ref{eq:H}, on a square lattice with PBC, which can be seen to have a single-particle dispersion relation
\begin{equation}
	E(k_x,k_y)
	= \sqrt{
	\left(\mu-2\sum_{i=x,y} \cos k_i \right)^2
	+ \left(2\gamma \sum_{i=x,y} \sin k_i \right)^2}.
\end{equation}

We can now distinguish three types of systems, as a function of the couplings $(\mu,\gamma)$: (1) gapped systems, where the single-particle energy $E(k_x,k_y)$ does not vanish for any $(k_x,k_y)$; (2) gapless systems with only a finite number of points $(k_x,k_y)$ where the energy $E(k_x,k_y)$ vanishes, that is, with a zero-dimensional Fermi surface; and (3) gapless systems with a continuum of points $(k_x,k_y)$ where the energy $E(k_x,k_y)$ vanishes, that is, with a one-dimensional Fermi surface.

\begin{figure}[h]
\begin{center}
\includegraphics[scale=0.7]{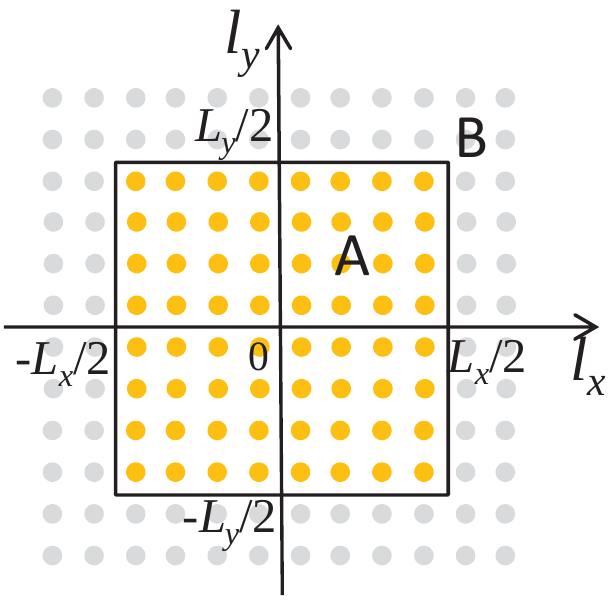}
\caption{
\label{fig:2dsetting}
Square lattice made of $N=N_x\times N_y$ sites, with a square region $A$ made of $L=L_x\times L_y$ sites, with $L_x = L_y$. In the main text we consider the thermodynamic limit, $N_x = N_y = \infty$. The coordinate of the sites are $(l_x,l_y)$, with $l_x, l_y = -\frac{L_x}{2}+\frac{1}{2}, ..., \frac{L_x}{2}-\frac{1}{2}$. The perimeter of the block is then proportional to $L_x$.
}
\end{center}
\end{figure}

To study the entanglement contour in these three types of systems, we consider an infinite square lattice and a square region $A$ made of $L = L_x\times L_y$ sites, for $L_x=L_y$, according to Fig. \ref{fig:2dsetting}.

\subsection{Gapped Hamiltonians}

\begin{figure}[h]
\begin{center}
\includegraphics[scale=0.45]{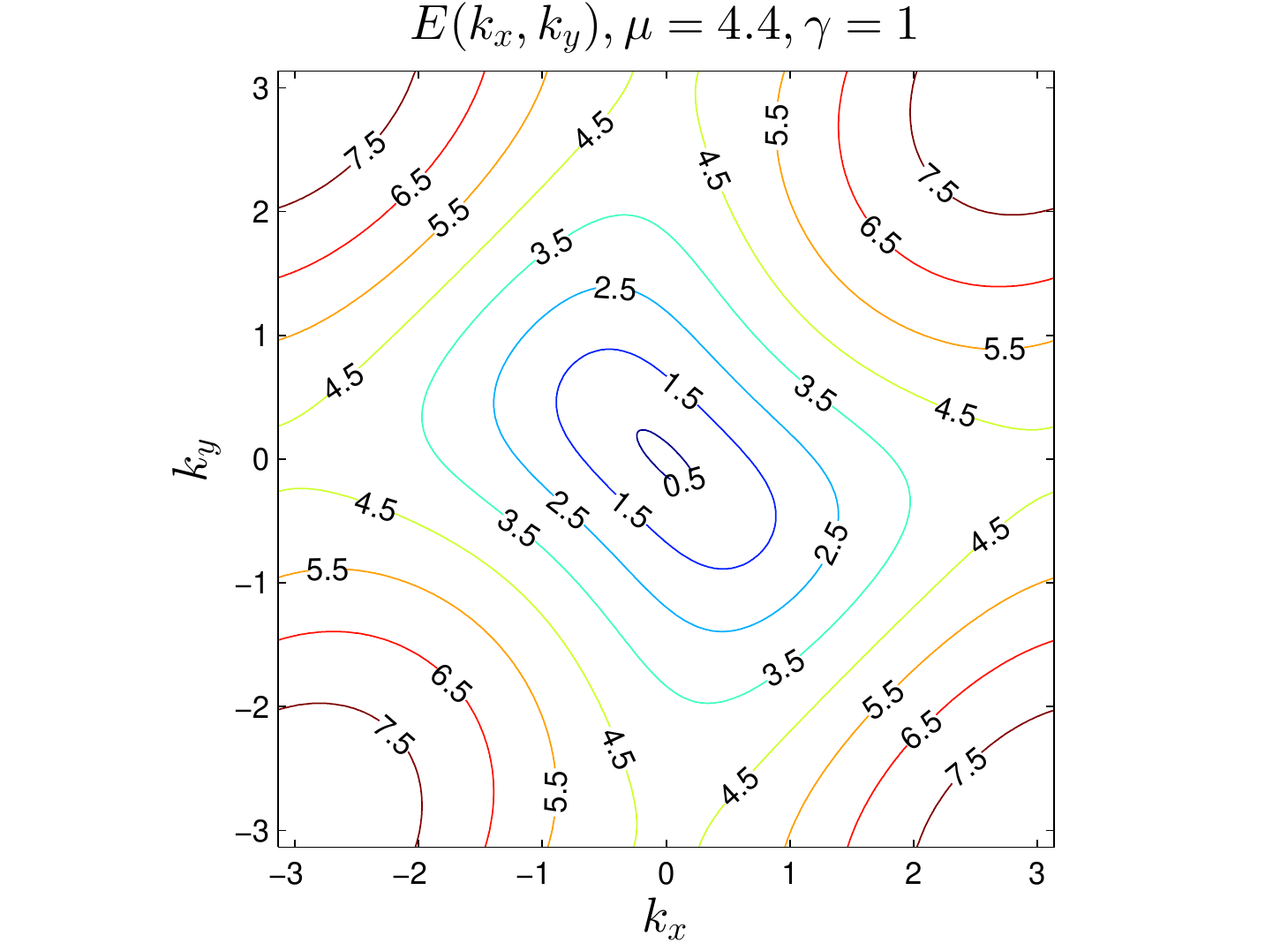}
\includegraphics[scale=0.5]{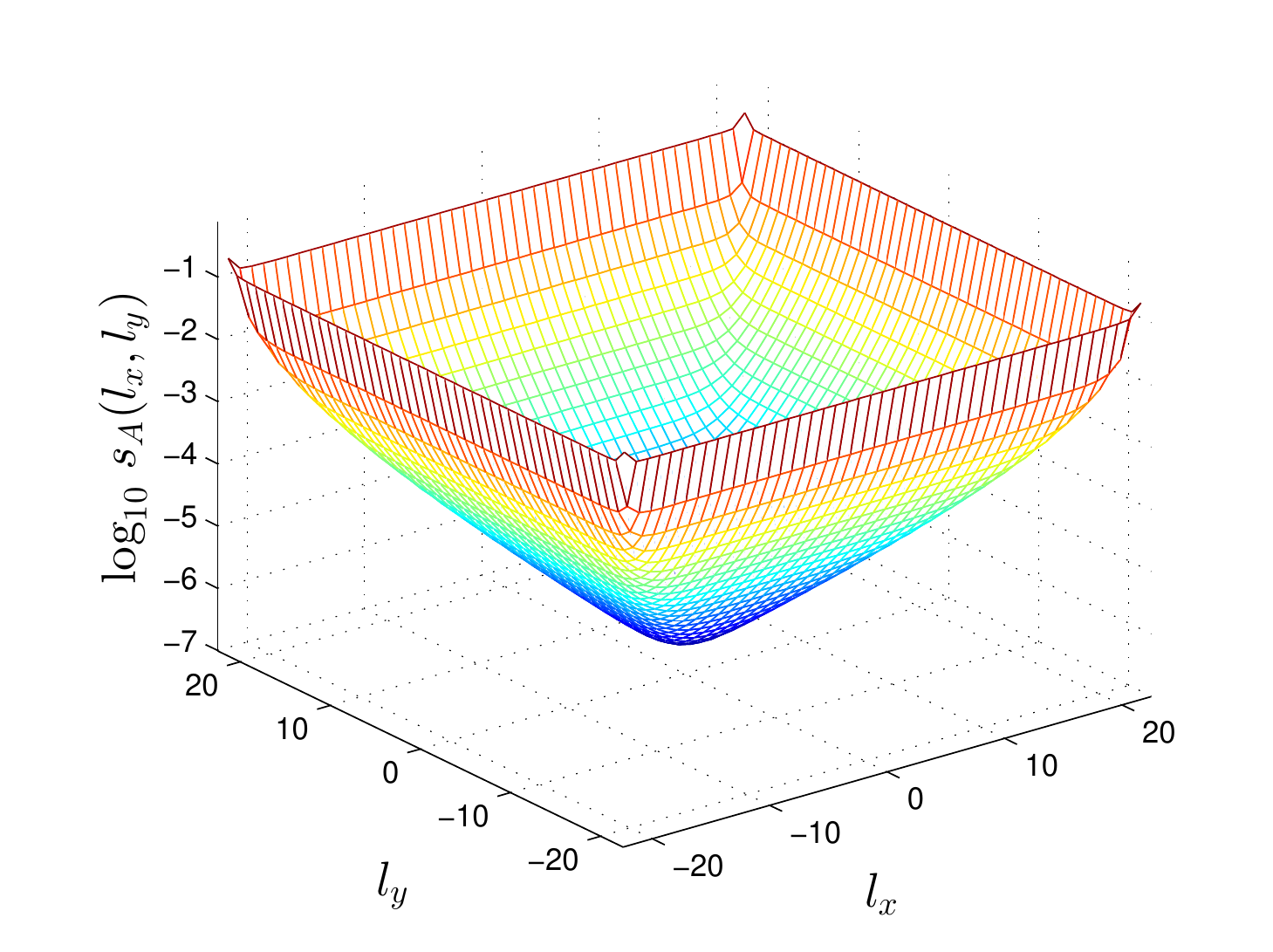}
\caption{
\label{fig:2dContour1}
Entanglement contour of a square region $A$ in a gapped fermionic system, $(\mu = 4.4, \gamma = 1)$, in $D=2$ dimensions. (Top:) The dispersion relation $E(k_x,k_y)$ has no zero-modes. (Bottom): Entanglement contour $s_A(l_x,l_y)$ of the squared region $A$, for $L_x = L_y = 40$. The cross section $s_A(l_x,0)$, shown in Fig. \ref{fig:2dContourX}, decays roughly exponentially with the distance to the boundary of region $A$.
}
\end{center}
\end{figure}

In a gapped system, the entanglement entropy is known to obey a boundary law, Eq. \ref{eq:boundary}. For a square region $A$ of perimeter $4L_x$ this reads
\begin{equation}
	S(A) = \alpha L_x + O(1),
\end{equation}
where the coefficient $\alpha$ is non-universal, whereas the constant term may reflect geometric properties of region $A$, as well as the presence of topological order in the ground state. Again, inspired by the exponential decay of the off-diagonal elements of $\GGamma^A$, we expect a contour that decays exponentially with the distance to the boundary of $A$. Its summation over the whole region $A$ would then immediately lead to the boundary law, as it did in $D=1$ dimensions. Fig. \ref{fig:2dContour1} shows the entanglement contour $\sss$ for ($\mu=4.4, \gamma = 1$), which corresponds to a gapped dispersion relation. We see that the contour decays roughly exponentially with the distance to the boundary, as expected.

\subsection{Gapless Hamiltonians with a finite number of zero modes}

\begin{figure}[h]
\begin{center}
\includegraphics[scale=0.45]{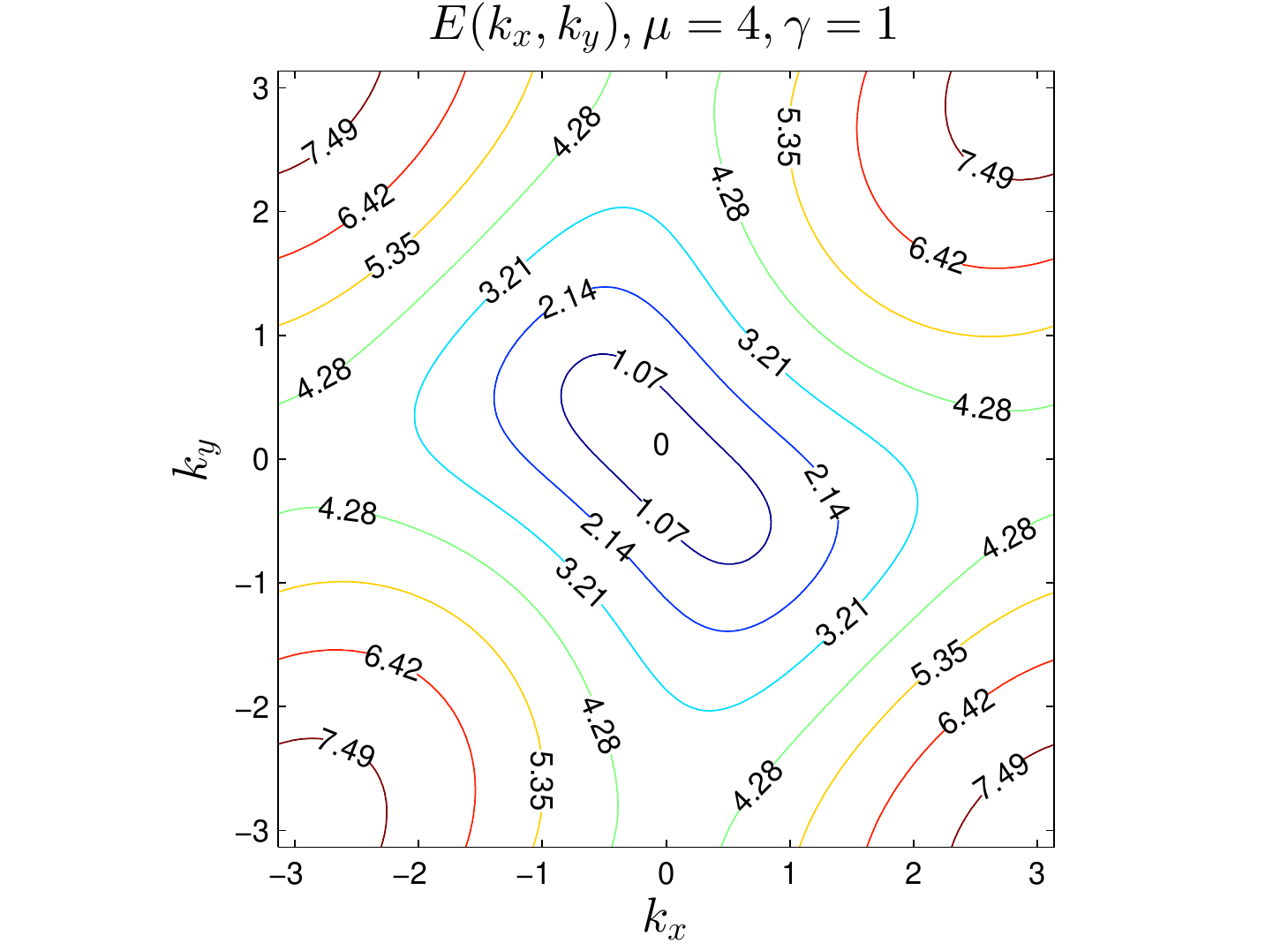}
\includegraphics[scale=0.5]{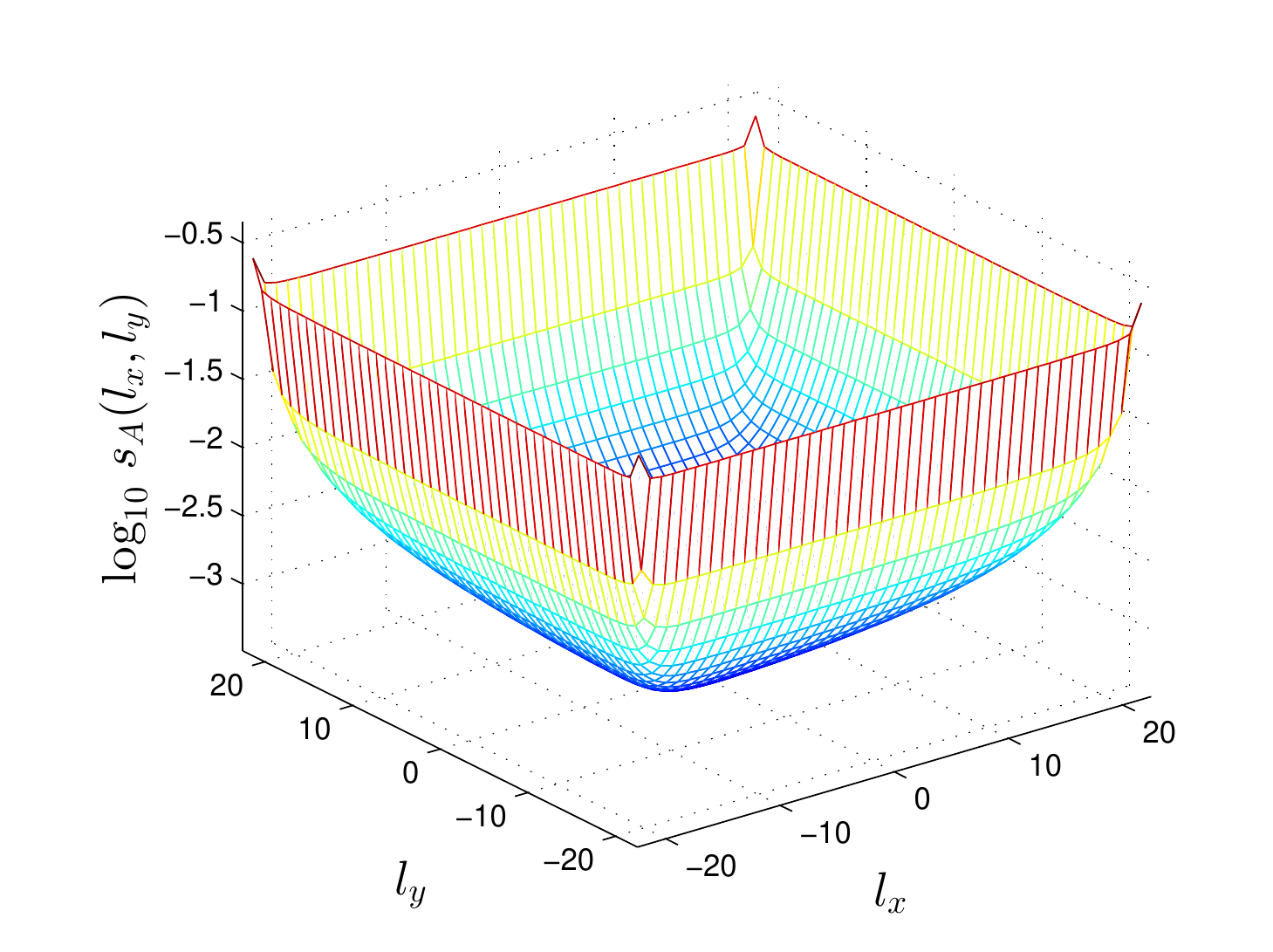}
\caption{
\label{fig:2dContour2}
Entanglement contour of a square region $A$ in a gapless fermionic system with a finite number of zero modes, $(\mu = 4, \gamma = 1)$, in $D=2$ dimensions. (Top:) The dispersion relation $E(k_x,k_y)$ only vanishes at $(k_x=0,k_y=0)$. (Bottom): Entanglement contour $s_A(l_x,l_y)$ of the squared region $A$, for $L_x = L_y = 40$. The cross section $s_A(l_x,0)$, shown in Fig. \ref{fig:2dContourX}, decays slower than in the gapped system. Notice also that at the center of the square ($l_x=l_y=0$), where the contour is smallest, it is several orders of magnitude larger than in the gapped case, Fig. \ref{fig:2dContour1}
}
\end{center}
\end{figure}

In a gapless system with a finite number of zero modes, the entanglement entropy is known to again obey the boundary law, Eq. \ref{eq:boundary}. That is, the scaling of the entanglement entropy is not sufficient in order to distinguish a gapped system from a gapless system. However, since the off-diagonal elements in $\GGamma^A$ have now a power-law decay, we expect that the entanglement contour will allow us to differentiate gapped and gapless systems. Fig. \ref{fig:2dContour2} shows the entanglement contour $\sss$ for ($\mu=4, \gamma = 1$), which has one zero mode in the dispersion relation. We see that in this case the contour decays slower than exponentially, thus allowing us to use the contour to differentiate this system from a gapped system. 

Similar conclusions can be obtained using other measures of entanglement. For instance, in the third paper of Ref. \cite{MutualInfo} an algebraic (and therefore slower than exponential) decay of the mutual information in a gapless system was reported. 

\subsection{Gapless Hamiltonians with a continuum of zero modes}

\begin{figure}[h]
\begin{center}
\includegraphics[scale=0.45]{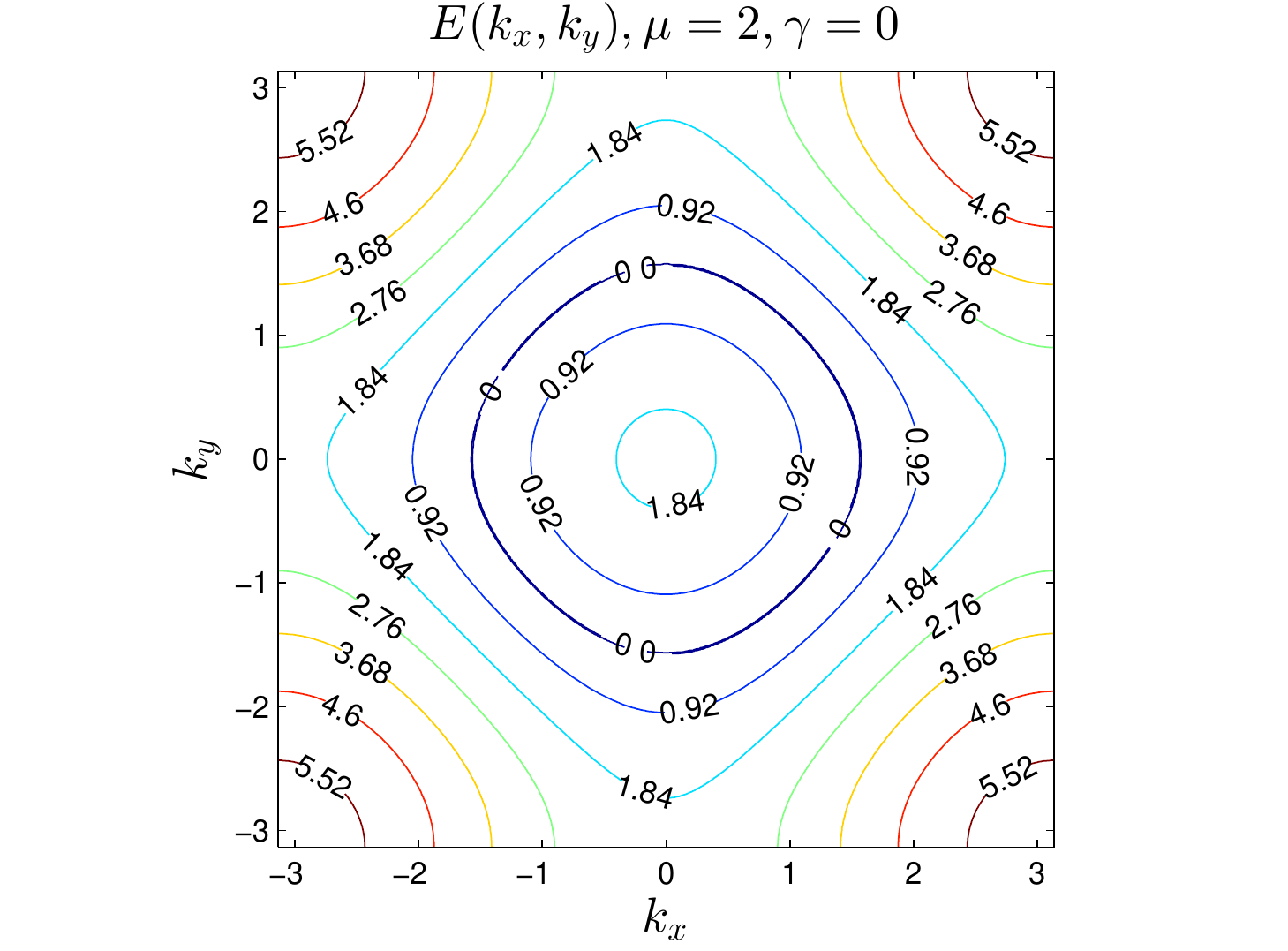}
\includegraphics[scale=0.5]{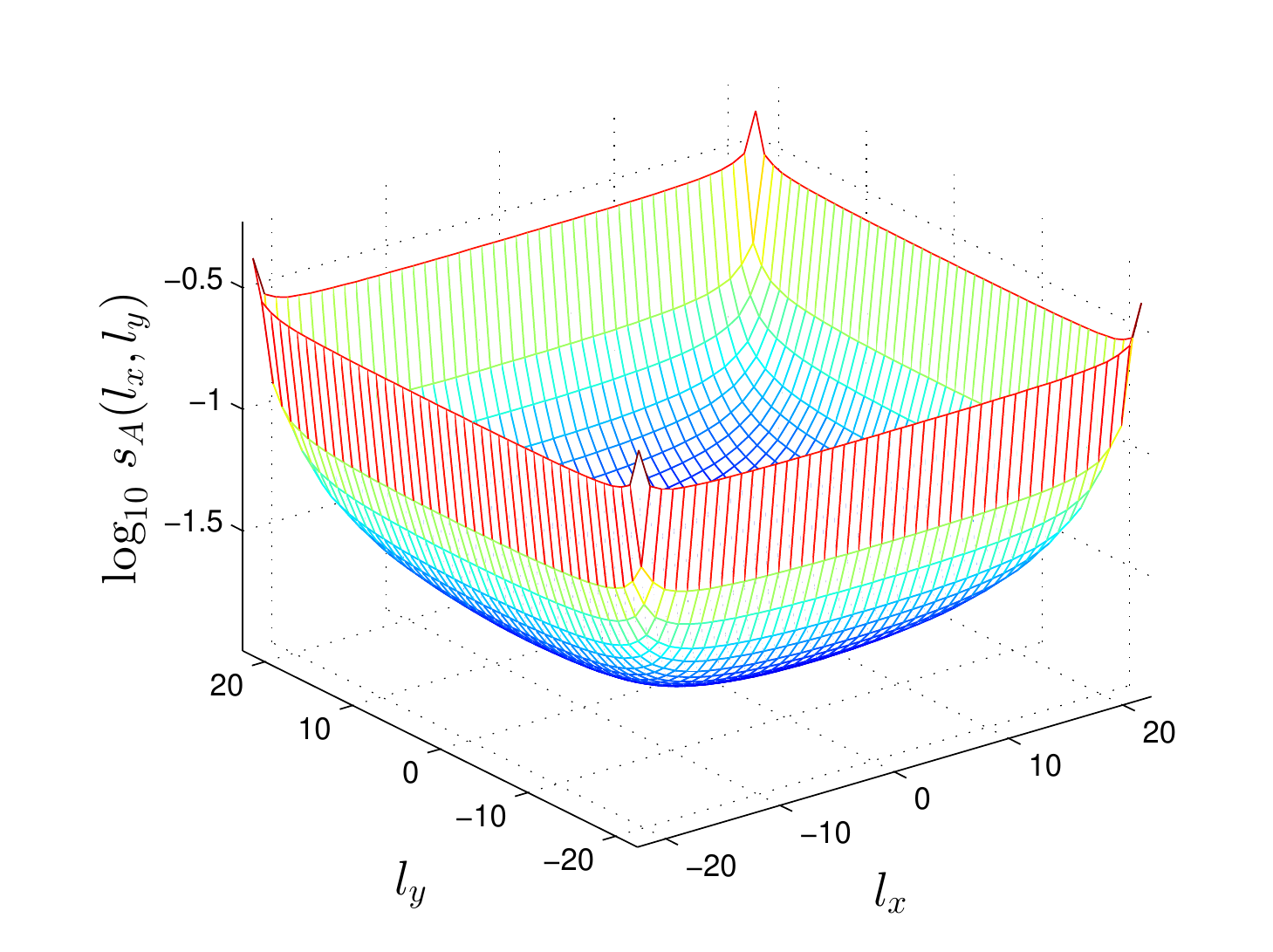}
\caption{
\label{fig:2dContour3}
Entanglement contour of a square region $A$ in a gapless fermionic system with a continuum of zero modes, $(\mu = 2, \gamma = 0)$, in $D=2$ dimensions. (Top:) The dispersion relation $E(k_x,k_y)$ vanishes for any solution of $\cos(k_x)+\cos(k_y)=1$. (Bottom): Entanglement contour $s_A(l_x,l_y)$ of the squared region $A$, for $L_x = L_y = 40$. The cross section $s_A(l_x,0)$, shown in Fig. \ref{fig:2dContourX}, decays even slower than in the gapless system with just one zero-mode.  At the center of the square ($l_x=l_y=0$), where the contour is smallest, it is still more than one order of magnitude larger than in Fig. \ref{fig:2dContour2}.
}
\end{center}
\end{figure}

\begin{figure}[h]
\begin{center}
\includegraphics[scale=0.5]{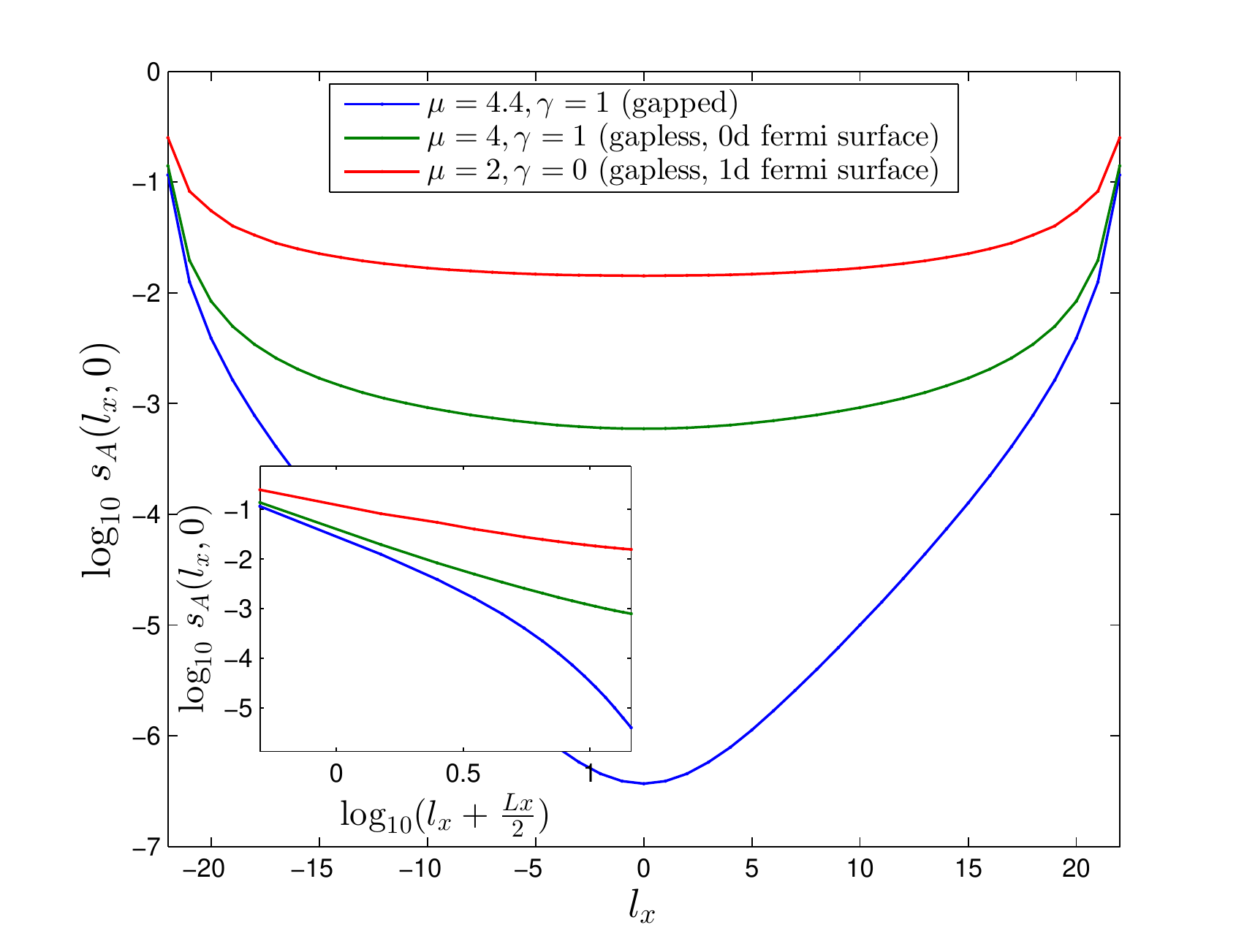}
\caption{
\label{fig:2dContourX}
Comparison of the cross section $s_A(l_x,0)$ of the contours under different parameters $(\mu,\gamma)$ of the Hamiltonian, corresponding to the gapped and gapless systems described in Figs. \ref{fig:2dContour1}, \ref{fig:2dContour2}, and \ref{fig:2dContour3}. Notice that $s_A(l_x,0)$ decays roughly exponentially with the distance to the boundary in the gapped system, whereas it decays significantly more slowly in the gapless systems. The inset shows the cross section $s_A(l_x,0)$ near the left boundary (in log-log scale). In the gapless case with one zero-mode, the contour seems to decay as a power-law with the distance to the boundary.
}
\end{center}
\end{figure}

Finally, in a gapless system with a continuum of zero modes, the entanglement entropy is known to display a logarithmic correction to the boundary law, Eq. \ref{eq:logarithm}, namely
\begin{equation}
	S(A) = \alpha L_x \log_2 L_x + \cdots,
\end{equation}
where $\alpha$ is again some non-universal constant and where we have only specified the leading contribution. In this case, the logarithmic correction already allows us to use the entanglement entropy to diagnose the presence of the Fermi surface. Fig. \ref{fig:2dContour3} shows the entanglement contour $\sss$ for ($\mu=2, \gamma = 0$), which has a continuum of zero modes corresponding to the solutions $(k_x,k_y)$ of $\cos(k_x)+\cos(k_y)=1$. We see that the contour decays even slower than in the presence of only a finite number of zero modes.

Fig. ~\ref{fig:2dContourX} compares the decay of the contour, by focusing on the section $s_A(l_x,0)$, for the three different types of systems discussed above.

\section{Time evolution after a quench}
\label{sect:quench}
In the previous two sections we have investigated the entanglement contour on the ground state of a local Hamiltonian $H$. In this section we explore what happens to the entanglement contour of the ground state under a sudden change of the Hamiltonian $H$, which is replaced by a new Hamiltonian $H'$, generating a non-trivial time evolution.

We will investigate both a local quench, in which $H$ and $H'$ differ only on a localized region (in our case, the Hamiltonian term linking regions $A$ and $B$), and a global quench, where $H$ and $H'$ differ everywhere in the lattice. Refs. \cite{LocalQuench} and \cite{GlobalQuench} contain examples of previous work in which a number of other tools have been used to characterize the evolution of entanglement after a local and global quench, respectively, leading to similar insight as the one obtained here with the entanglement contour.

\subsection{Local quench}

\begin{figure}[h]
\begin{center}
\includegraphics[scale=0.7]{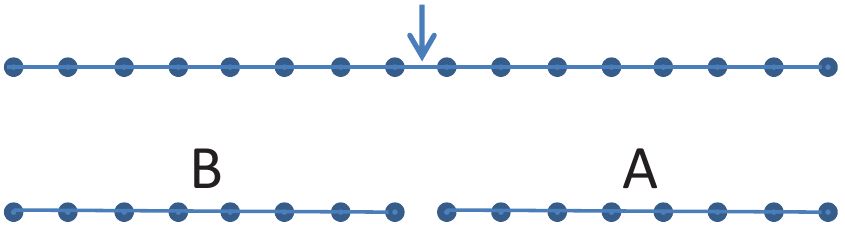}
\caption{
\label{fig:localquench}
Local quench on a $D=1$ dimensional lattice with open boundary conditions. The Hamiltonian term acting across the bond that connects regions $A$ and $B$ is removed in the new Hamiltonian $H' = H_A + H_B$. As a result, regions $A$ and $B$ evolve independently according to the time evolution operator $e^{-iH_At}e^{-iH_Bt}$.
}
\end{center}
\end{figure}

Let us consider a fermionic lattice model in $D=1$ dimensions with a quadratic Hamiltonian $H$ given by Eq. \ref{eq:Hamiltonian} with $(\mu=3,\gamma=1)$, which corresponds to a gapped system. We use the setting of Fig. \ref{fig:1Dchain1}, with open boundary conditions and a region $A$ corresponding to the right half of the lattice. At time $t=0$ the system is in its ground state. The entanglement entropy $S(A,t=0)$ is some finite constant and its contour decays exponentially with the distance to the boundary cut between regions $A$ and $B$, which is the left half of the chain. The contour $\sss(l)$ at $t=0$ is shown in the upper panel of Fig. \ref{fig:localquench} (left-most line, in blue).

\begin{figure}[h]
\begin{center}
\includegraphics[scale=0.55]{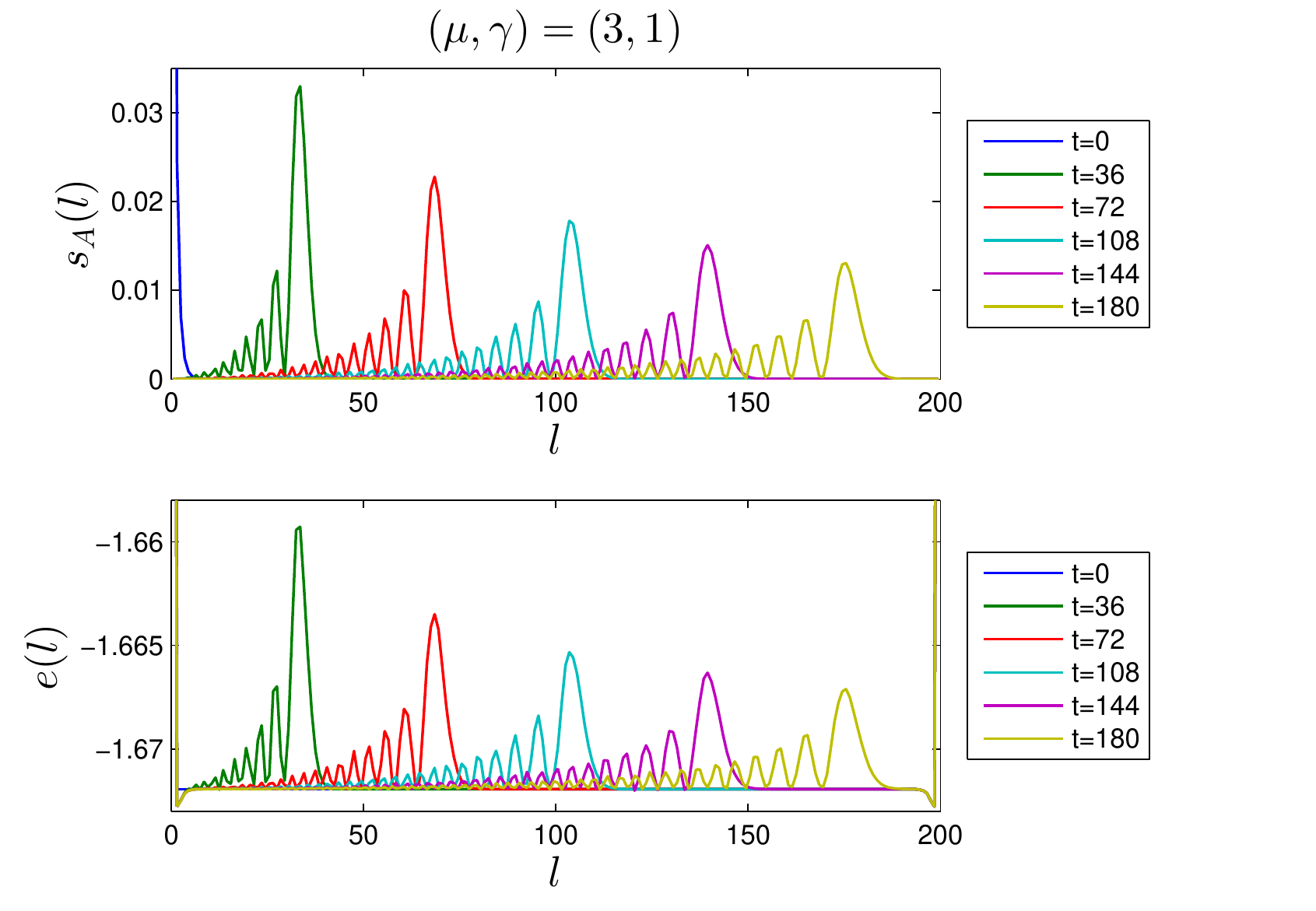}
\includegraphics[scale=0.45]{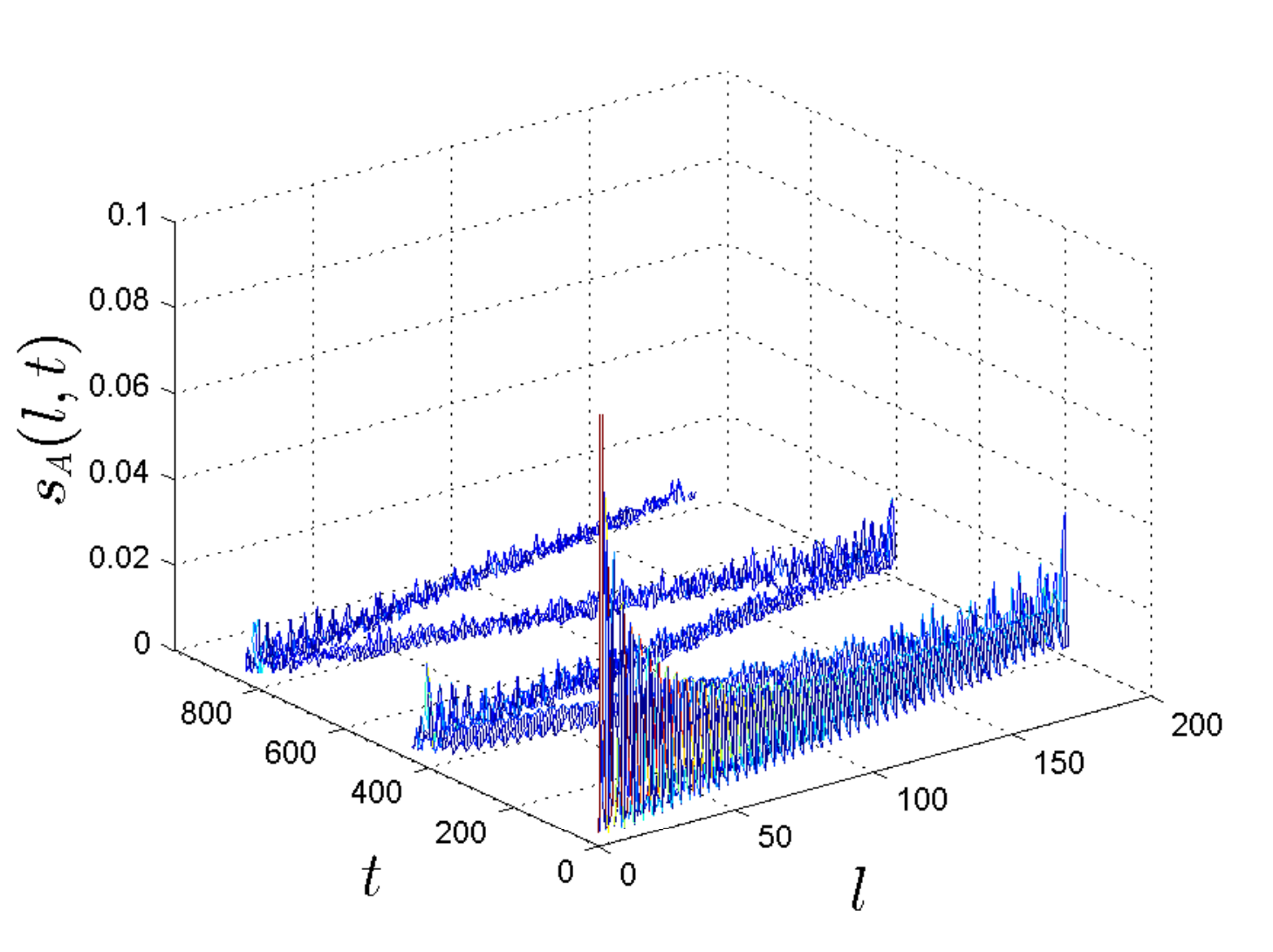}
\caption{
\label{fig:LocalQuench1}
Entanglement contour after a local quench in a lattice made of $N=400$ sites, with open boundary conditions and a gapped Hamiltonian $(\mu,\gamma)=(3,1)$. Region $A$ is made of $L=200$ sites. (Top:) $s_A(l,t)$ for different values of $t$. The contour shows that the subregion within region $A$ that is entangled with region $B$ changes in time, progressing from left to right. (Middle:) Interestingly, the energy density describes a very similar behavior. (Bottom:) A three dimensional plot of $s_A(l,t)$, with the peak of the wave highlighted, showing that when the entanglement wave reaches the end of region $A$ it bounces back.
}
\end{center}
\end{figure}

Then, from $t=0$ onwards, the system evolves unitarily according to a new Hamiltonian $H'=H_A + H_B$ that differs from $H$ in that the hopping term connecting regions $A$ and $B$ has been removed. Notice that, as a result, the time evolution operator $e^{-iH't} = e^{-i (H_A+H_B) t}$ factorizes as the product of two time evolution operators $e^{-iH_At}$ and $e^{-iH_Bt}$ for the two regions $A$ and $B$. It follows that no entanglement between these two regions is produced or destroyed during the time evolution, and therefore $S(A,t) = S(A,0)$ for all times $t > 0$. However, a constant entanglement entropy for region $A$ does not imply that its contour is also constant. On the contrary, as shown in Fig. \ref{fig:LocalQuench1}, the contour $\sss(l,t)$ has a distinct evolution in time, spreading through region $A$ as a wave (or, rather, a tsunami), indicating that the degrees of freedom in $A$ that are entangled with $B$ are no longer located near the boundary with $B$.

Notice that, under time evolution, the entanglement contour behaves as if it was a density of some locally conserved quantity -- and, as a matter of fact, in the example we study it follows closely the behavior of the energy density, shown in the middle panel of Fig. \ref{fig:LocalQuench1} for comparison. All these observations seem compatible with the existence of propagating pseudo-particles as proposed in earlier work and analyzed using correlations and entanglement entropies for a multiplicity of cuts \cite{LocalQuench}.

\subsection{Global quench}

\begin{figure}[h]
\begin{center}
\includegraphics[scale=0.48]{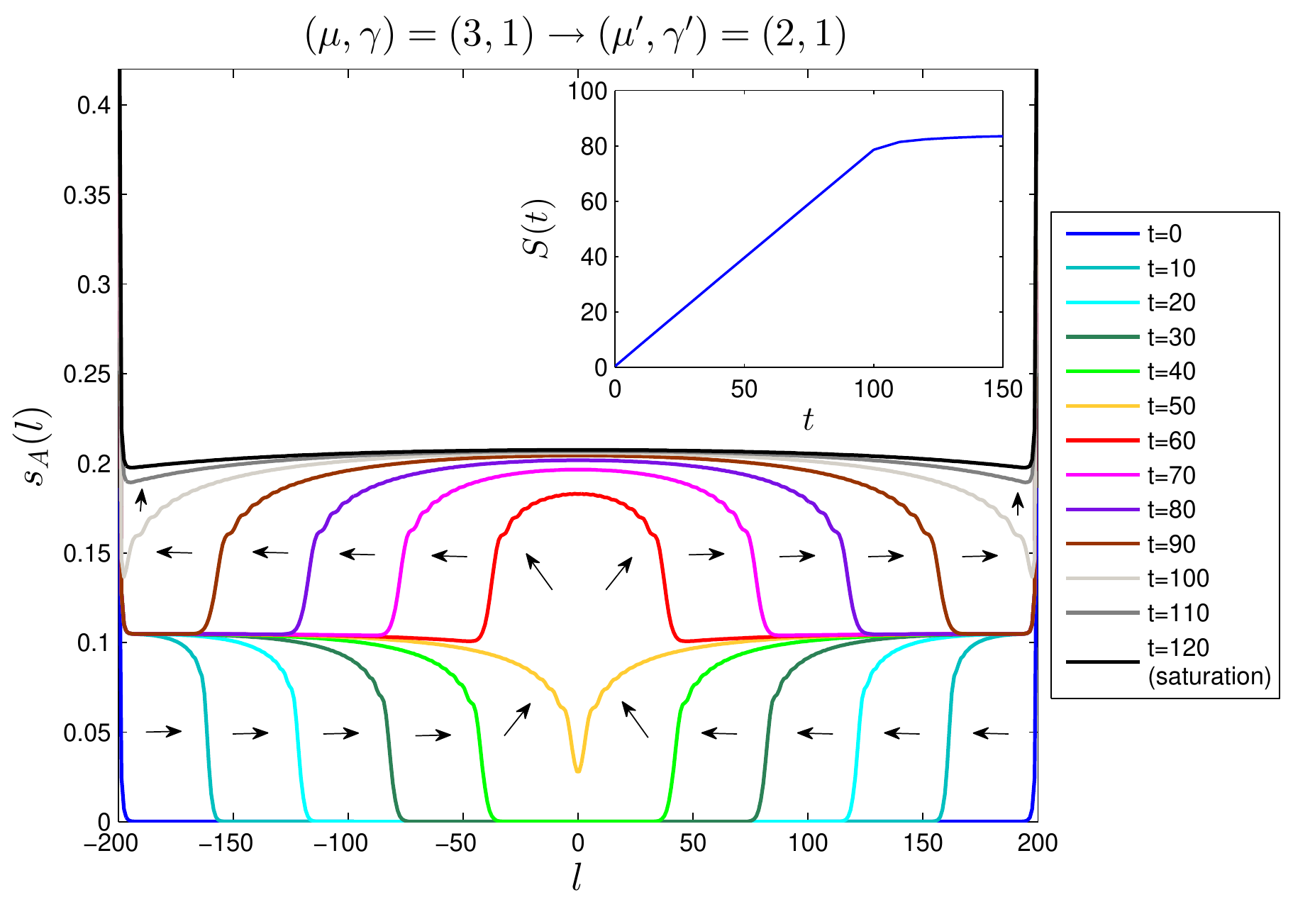}
\caption{
\label{fig:GlobalQuench1}
Entanglement contour after a global quench in a lattice made of $N=\infty$ sites. The ground state of a gapped Hamiltonian $H$ with $(\mu,\gamma)=(3,1)$ is made evolve according to a gapless Hamiltonian $H'$ with $(\mu',\gamma')=(2,1)$. The figure shows the entanglement contour $s_A(l,t)$ of a region $A$ made of $L=400$ sites. An entanglement front progresses from left to right (and from right to left), as described in the main text. The inset shows the entanglement entropy $S(A,t)$, which grows linearly in time while the two entanglement fronts progress front one boundary of $A$ to the opposite one and then back, and which saturates to a constant once the entanglement fronts have completed their progression through region $A$
}
\end{center}
\end{figure}

Let us now consider the entanglement contour after a global quench.
As in the previous example, we consider a $D=1$ dimensional lattice with the quadratic Hamiltonian $H$ given by Eq. \ref{eq:Hamiltonian} with $(\mu=3,\gamma=1)$, which corresponds to a gapped system. This time, however, we choose the setting of Fig. \ref{fig:1Dchain2}, namely an infinite lattice with a region $A$ of finite size $L$ and with  two boundary cuts. At time $t=0$ the system is in its ground state. The entanglement entropy $S(A,t=0)$ is some finite constant and its contour decays exponentially with the distance to each of the two boundary cuts. The contour $\sss(l)$ at $t=0$ can be seen in Fig. \ref{fig:GlobalQuench1} (left-most and right-most lines at the bottom, in blue).

From $t=0$ onwards, the system evolves according to the unitary time evolution operator $e^{-iH't}$, where the Hamiltonian $H'$ corresponds to Eq. \ref{eq:Hamiltonian} with $(\mu=2,\gamma=1)$. That is, we have made a sudden change in the chemical potential, such that the system has become gapless. Under this global quench, regions $A$ and $B$ are still connected by $H'$, and therefore entanglement entropy can be created during the time evolution. As shown in the inset of Fig. \ref{fig:GlobalQuench1}, $S(A,t)$ grows linearly as a function of $t$, starting from an almost vanishing value at $t=0$, until it reaches some form of saturation at time $t \approx 100-120$.

The entanglement contour offers again a much more detailed description of the time-dependent, spatial structure of quantum entanglement. As shown in Fig. \ref{fig:GlobalQuench1}, we can see that the degrees of freedom in $A$ that are entangled with $B$ are initially concentrated only near the boundary cuts. However, as time passes, more degrees of freedom in region $A$, further away from the boundary cuts, also contribute to the entanglement with region $B$. We can see two \textit{entanglement fronts} progressing at constant speed from the two boundaries towards the center of region $A$, for times $t\approx [0, 50-60]$. The two entanglement fronts cross without interfering with each other (but since the contour is the sum of the contributions by the two fronts, their crossing results in a new plateau that now grows from the center of region $A$) and then continue to advance until they hit the boundary of $A$ opposite to the one they started from, at time $t\approx 100-120$). Then the contour $\sss(l,t)$ has become roughly flat and stops evolving in time. This coincides in time with the saturation of the entanglement entropy $S(A,t)$.

In summary, the linear regime in the growth of $S(A,t)$ in time is explained in terms of the constant speed of the two entanglement fronts, whereas the saturation of $S(A,t)$ at long times is a consequence of the arrival of the entanglement fronts to the opposite ends of region $A$.

Again, the above observations appear compatible with the existence of propagating pseudo-particles, as analyzed using e.g. mutual information \cite{GlobalQuench} or entanglement density \cite{Takayanagi}.

\section{Discussion and outlook}
\label{sect:discussion}

In this paper we have introduced the notion of entanglement contour $s_A$ as a means to characterize which degrees of freedom of a region $A$ contribute to the entanglement entropy $S(A)$ between region $A$ and the rest $B$ of an extended quantum many-body system. We have proposed several properties that the entanglement contour should satisfy, and have identified a quantity $\sss$ for fermionic gaussian states which fulfills these properties.

Then we have explored the entanglement contour in selected ground states and during the time evolution that ensues a local or global quantum quench. This has allowed us to illustrate, through concrete examples, the potential of the entanglement contour as a means to provide a characterization of the entanglement between regions $A$ and $B$ than is finer than the one provided by the entanglement entropy $S(A)$ alone.

We have been able to formalize, and quantitatively characterize, the intuition that ground state entanglement in gapped systems involves, essentially, only the degrees of freedom near the boundary of region $A$. Moreover, we have seen that, in gapless systems, also degrees of freedom that are far away from the boundary of $A$ contribute significantly to the entanglement entropy $S(A)$. In $D=1$ dimensions the entanglement contour $s_A$ has allowed us to estimate the central charge $c$ of the underlying CFT by studying a single region $A$ (as opposed to multiple regions with different sizes, as needed in order to extract the central charge from the entanglement entropy $S(A)$); and in $D=2$ dimensions it has allowed us to discriminate between gapped systems and gapless systems with a finite number of zero modes (these two type of system obey a boundary law for the entanglement entropy $S(A)$, and therefore cannot be distinguished using only the scaling of $S(A)$).

Finally, we have seen that the entanglement contour reveals a detailed real-space structure of the entanglement of a region $A$ and its dynamics, well beyond what is accessible from the entanglement entropy $S(A)$ alone. For instance, after a local quench that detaches region $A$ from region $B$, we have seen that while $S(A)$ remains constant, the entanglement propagates within region $A$ in a wave-like manner, suggesting that perhaps its dynamics obeys a simple differential equation. On the other hand, after a global quench, we have seen an entanglement front sweeping region $A$ at constant speed from each boundary to the opposite one, offering insights into the well-known linear growth, then saturation of the entanglement entropy $S(A,t)$.

Most of the results presented in this paper can be generalized to free fermionic quantum fields, and also to lattice models and quantum field theories of free bosons. Free fermion/boson models, on the lattice or in the continuum, have played and continue to play a leading role in the development of a theory of entanglement in extended quantum systems. We hope that the entanglement contour $\sss$ will add to the previously existing tools used to characterize entanglement in such systems. Needless to say, however, the truly interesting (and much more challenging) extended systems are those of interacting particles. We leave the study of an entanglement contour for interacting systems for future work.

The authors thank Robert Myers for extensive discussions, and Chung Ming-Chiang, Tadashi Takayanagi, Brian Swingle, and an anonymous referee for valuable comments and point out existing literature. 
G.V. acknowledges support by NSERC (discovery grant), the John Templeton Foundation and by the Simons Foundation (Many Electron Collaboration).
Research at Perimeter Institute is supported by the Government of Canada through Industry Canada and by the Province of Ontario through the Ministry of Research and Innovation.


\end{document}